\documentclass{MYappolb}
\usepackage{graphicx}
\usepackage{cite}
\usepackage{amsfonts}\usepackage{amssymb,amsmath}
\usepackage{epstopdf}
\def\be{\begin{equation}}
\def\ee{\end{equation}}
\def\beqn{\begin{eqnarray}}
\def\eeqn{\end{eqnarray}}
\def\no{\nonumber}
\def\ba{\begin{array}{c}}
\def\bat{\begin{array}{cc}}
\def\ea{\end{array}}
\def\bi{\begin{itemize}}
\def\ei{\end{itemize}}

\def\cL{{\cal L}}

\def\cC{{\cal C}}

\def\cO{{\cal O}}
\def\cR{{\cal R}}
\def\cY{{\cal Y}}
\def\cZ{{\cal Z}}
\def\cF{{\cal F}}
\newcommand{\eqn}[1]{(\ref{#1})}
\newcommand{\bel}[1]{\be\label{#1}}
\newcommand{\rms}{\rm\scriptsize}
\def\gp{g\hskip .7pt\raisebox{-1.5pt}{$'$}}

\begin{document}
\title{Electroweak Symmetry Breaking and the Higgs Boson%Physics
\thanks{55th Cracow School of Theoretical Physics, Zakopane, Poland, 20--28 June 2015}
% \\ Particles and resonances of the Standard Model and beyond}%
}

\author{Antonio Pich
\address{Departament de F\'{\i}sica Te\`orica, IFIC, Universitat de Val\`encia -- CSIC\\
Apartat Correus 22085, E-46071 Val\`encia, Spain}
}
\maketitle
\begin{abstract}
The first LHC run has confirmed the Standard Model as the correct theory at the electroweak scale, and the existence of a Higgs-like particle associated with the spontaneous breaking of the electroweak gauge symmetry. These lectures overview the present knowledge on the Higgs boson and discuss alternative scenarios of electroweak symmetry breaking which are already being constrained by the experimental data.

\end{abstract}
% 12.60.-i, 14.80.-j,

\section{Introduction}

The LHC experiments ATLAS \cite{Aad:2012tfa} and CMS \cite{Chatrchyan:2012xdj} discovered in 2012 a massive state $H$ with the properties expected for a (Brout-Englert-Guralnik-Hagen-Kibble)-Higgs boson \cite{Higgs:1964pj,Higgs:1964ia,Higgs:1966ev,Englert:1964et,Guralnik:1964eu,Kibble:1967sv}.
The simple observation of the $H\to 2\gamma$ decay mode already demonstrated some basic characteristics of the new particle: it is electrically neutral, colourless and of integer spin, {\it i.e.}, a boson; moreover, conservation of angular momentum plus Bose symmetry imply that
$J\not=1$ \cite{Landau:1948kw,Yang:1950rg}. The angular distributions of the final lepton pairs in $H\to Z Z^*\to\ell^-\ell^+\ell'^-\ell'^+$ decays \cite{Chatrchyan:2012jja,Aad:2013xqa} confirm the $J^P=0^+$ assignment; the $J^P=0^-$ and $2^+$ hypotheses being excluded at confidence levels above 99\%.
The masses measured by the two experiments are in good agreement, giving the average value \cite{Aad:2015zhl}
\begin{equation}\label{eq:Higgs_Mass}
M_H\, =\, (125.09 \pm 0.21\pm 0.11)~\mathrm{GeV}\, =\, (125.09 \pm 0.24)~\mathrm{GeV}\, .
\end{equation}

All data accumulated so far confirm the Standard Model (SM) as the appropriate theoretical description of the electroweak and strong interactions at the energy scales explored until now \cite{Pich:2015tqa}. The SM successfully explains the experimental results with high precision and all its ingredients, including the Higgs boson, have been finally verified.
An important question to be addressed is whether $H$ corresponds to the unique Higgs boson incorporated in the SM, or it is just the first signal of a much richer scenario
of Electroweak Symmetry Breaking (EWSB). Obvious possibilities are an extended scalar sector with additional fields or dynamical (non-perturbative) EWSB generated by some new underlying dynamics.
While more experimental analyses are needed to assess the actual nature of the $H$ boson, the
present data give already very important clues, constraining its couplings in a quite significant way.

Whatever the answer turns out to be, the LHC findings represent a truly fundamental discovery with far reaching implications. If $H$ is an elementary scalar (the first one), one would have established the existence of a bosonic field (interaction) which is not a gauge force. If  instead, it is a composite object, a completely new underlying interaction should exist.

The following sections contain an introduction to the EWSB and the physics of the Higgs boson.
The SM mechanism of EWSB is briefly reviewed in section~\ref{sec:SM}. Section~\ref{sec:Higgs-data} describes the current experimental knowledge on the Higgs properties. Quantum corrections and the important role played by the heavy top mass scale are discussed in sections~\ref{sec:loops} and \ref{sec:top}. Section~\ref{sec:xSM} analyzes the simplest extension of the SM scalar sector with a singlet scalar. The deep relation between flavour dynamics and EWSB is discussed in section~\ref{sec:mHDM} which considers models with several scalar doublets. Section~\ref{sec:custodial} discusses the custodial symmetry characterizing the SM EWSB and provides a very basic introduction to the electroweak effective theory. Some
% summarizing
comments on the present status are finally given in section~\ref{sec:outlook}.

\section{Standard Model Higgs Mechanism}
\label{sec:SM}

A massless gauge boson has only two possible polarizations, while a massive spin-1 particle should have three. To generate the missing longitudinal polarizations of the $W^\pm$ and $Z$ bosons, without breaking gauge invariance, one needs to incorporate three additional degrees of freedom to the $SU(2)_L\otimes U(1)_Y$ gauge Lagrangian \cite{Pich:2012sx}. The SM \cite{GL:61,WE:67,SA:69} adds a $SU(2)_L$ doublet of complex scalar fields
\begin{equation}
\Phi(x)\; \equiv\;\left(\ba \phi^{(+)}(x)\\ \phi^{(0)}(x)\ea\right)
\; =\; \exp{\left\{\frac{i}{v}\,\vec{\sigma}\cdot\vec{\varphi}(x)\right\}}\;\,\frac{1}{\sqrt{2}}\,
\left[\begin{array}{c} 0\\ v+H(x)\end{array}\right]\, ,
\label{eq:doublet}
\end{equation}
with a non-trivial potential generating the wanted EWSB:
%
%%%%%%%%%%%%%%%%%%%%%% Figure Higgs Potential %%%%%%%%%%%%%%%%%%%%%
\begin{figure}[htb]
%\centering
\begin{center}
\includegraphics[width=4.5cm,clip]{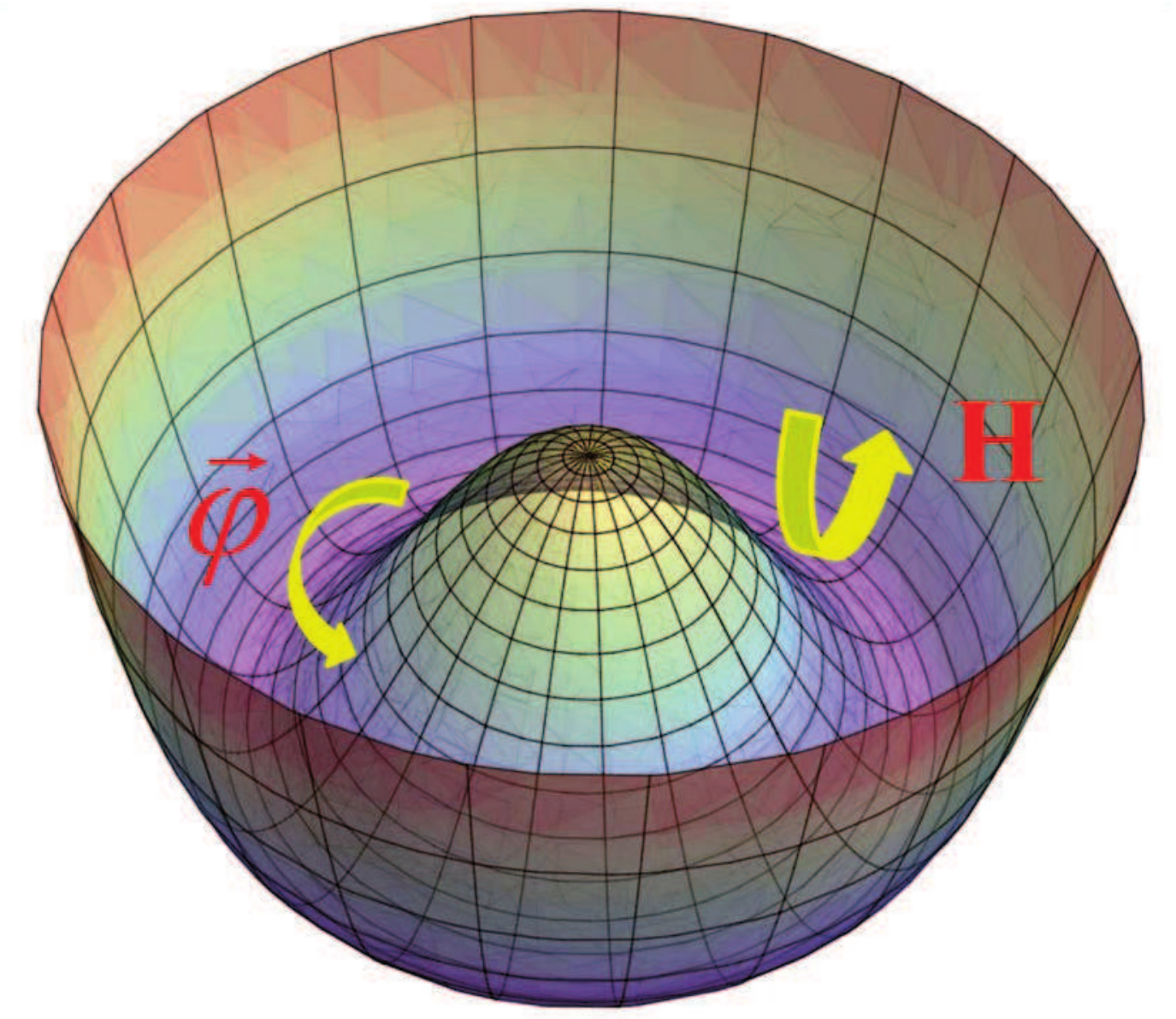}
\caption{SM scalar potential. The Goldstone ($\vec\varphi$) and Higgs ($H$) fields parametrize excitations along the directions indicated by the arrows.}
\label{fig:HiggsPotential}
\end{center}
\end{figure}
%%%%%%%%%%%%%%%%%%%%%%%%%%%%%%%%%%%%%%%%%%%%%%%%%%%%%%%%%%%%%%%%%%%
%
\begin{equation}\label{eq:Lphi}
\cL_\phi\; =\; (D^\mu\Phi)^\dagger D_\mu\Phi - \lambda\,\left( |\Phi|^2 -\frac{v^2}{2}\right)^2 .
% +\frac{\lambda}{4}\, v^4\, .
\end{equation}
In order to have a ground state the potential should be bounded from
below, {\it i.e.}, $\lambda >0$. The covariant derivative
\bel{eq:DS}
D_\mu\Phi \, = \, \left[\partial_\mu
+\frac{i}{2}\, g\;\vec{\sigma}\cdot\vec{W}_\mu
%\widetilde W^\mu
+ i\, \gp\, y_\phi\, B_\mu\right]\,\Phi\; ,
%\qquad\qquad
% y_\phi\, =\, Q_\phi - T_3 \, =\, \frac{1}{2}\, ,
\ee
couples the scalar doublet to the SM gauge bosons. The value of the scalar hypercharge,
$y_\phi = (Q - T_3)_\phi  = \frac{1}{2}$,
is fixed by the requirement of having the correct couplings between $\Phi(x)$ and $A^\mu(x)$: the photon should not couple to $\phi^{(0)}$, and $\phi^{(+)}$ must have the right electric charge. To preserve the conservation of the electric charge, only the neutral scalar field can acquire a vacuum expectation value.

As shown in Fig.~\ref{fig:HiggsPotential}, there is a infinite set of degenerate states with minimum energy, satisfying
$|\langle 0|\Phi|0\rangle | = {v/\sqrt{2}}$.
%%
%\bel{eq:vev}
%\big|\langle 0|\phi^{(0)}|0\rangle\big|
%\, = \, \sqrt{{-\mu^2\over 2 h}} \,\equiv\, {v\over\sqrt{2}} \, .
%\ee
%%
Once we choose a particular ground state, for instance $\vec\varphi = H = 0$ in Eq.~\eqn{eq:doublet},
the $SU(2)_L\otimes U(1)_Y$ symmetry gets spontaneously broken to
the electromagnetic subgroup $U(1)_{\rms QED}$, which by
construction still remains a true symmetry of the vacuum. According
to Goldstone's theorem \cite{NA:60,GO:61,GSW:62}, three massless states should then appear
(one for each broken generator of the symmetry group).
The Goldstone modes $\vec{\varphi}$ describe excitations along the flat directions of the potential, {\it i.e.}, into states with the same energy as the chosen ground state. Since those excitations do not cost any energy, they obviously correspond to massless states.

In the unitary gauge, $\vec\varphi(x) = \vec{0}$, the three Goldstone fields are removed and
the SM Lagrangian describes massive $W^\pm$ and $Z$ bosons; their masses being generated by the derivative term in Eq.~(\ref{eq:Lphi}). The scalar Lagrangian takes then the form:
\begin{equation}\label{eq:L_H}
\cL_\phi\; =\; \left(1+\frac{H}{v}\right)^2 \,
\left\{ M_W^2\, W_\mu^\dagger W^\mu
+ \frac{1}{2}\, M_Z^2\, Z_\mu Z^\mu \right\}\, +\, \cL_H\, ,
\end{equation}
with
\be
M_W\, =\, M_Z\, \cos{\theta_W}\, =\,\frac{1}{2}\, g\, v \, ,
\ee
where the weak mixing angle defining the $Z$ and $\gamma$ fields,
\bel{eq:Z_g_mixing}
\left(\ba W_\mu^3 \\ B_\mu\ea\right) \,\equiv\,
\left(\bat
\cos{\theta_W} & \sin{\theta_W} \\ -\sin{\theta_W} & \cos{\theta_W}
\ea\right) \, \left(\ba Z_\mu \\ A_\mu\ea\right)\, ,
\ee
is related to the gauge couplings through \
$g\,\sin{\theta_W} = \gp\,\cos{\theta_W} = e$.
The measured masses of the gauge bosons imply \
$\sin^2{\theta_W} = 1 - {M_W^2/M_Z^2}  = 0.223$.

The three Goldstones have been ``eaten up'' by the gauge bosons, giving rise to their longitudinal polarizations. The total number of degrees of freedom (dof) is of course the same.
A massive scalar field $H(x)$, the Higgs, remains in the physical spectrum of the electroweak theory because $\Phi(x)$ contains a fourth degree of freedom, which is not needed for the EWSB. Its couplings to the $W^\pm$ and $Z$ bosons, shown in Eq.~\eqn{eq:L_H}, are proportional to the square of their masses. The scalar potential generates the Higgs mass, and cubic and quartic self-interactions:
\be\label{eq:H_int}
\cL_H \; = \; {1\over 2}\, \partial_\mu H\, \partial^\mu H -
{1\over 2}\, M_H^2\, H^2
- {M_H^2\over 2 v}\, H^3 - {M_H^2\over 8 v^2}\, H^4\, .
\ee

The doublet structure of the complex scalar field $\Phi(x)$ provides a renormalizable model \cite{TH:71} with good unitarity properties.
While the vacuum expectation value (the electroweak scale) was already known from the $\mu^-\to e^-\bar\nu_e\nu_\mu$ decay rate,
\bel{eq:G_F}
v\, =\, (\sqrt{2}\, G_F)^{-1/2}\, =\, 246~\mathrm{GeV}\, ,
\ee
the measured Higgs mass determines the last free parameter of the SM, the quartic scalar coupling:
\begin{equation}\label{eq:lambda}
\lambda\; =\; \frac{M_H^2}{2 v^2}\; =\; 0.13\, .
\end{equation}

\subsection{Fermion Masses}

Fermionic mass terms are forbidden by the $SU(2)_L\otimes U(1)_Y$ symmetry because they
would mix the left and right-handed components of the fermion fields, which transform differently under the SM gauge group.
However, the additional scalar doublet allows us to write gauge-invariant fermion-scalar couplings:
\bel{eq:yukawa2}
\cL_Y\, =\, -y_d\, \bar Q_L \,\Phi\, d_R \, - \,
y_u\,\bar Q_L\, \tilde\Phi\, u_R \, - \,
y_\ell\,\bar L_L \,\Phi\, \ell_R
\, +\, \mbox{\rm h.c.}\, ,
\ee
where $\bar Q_L = (\bar u_L, \bar d_L)$ and $\bar L_L = (\bar \nu_L, \bar \ell_L)$ are the quark and lepton left-handed doublets (for a single family), $u_R$, $d_R$ and $\ell_R$ the corresponding right-handed fermion singlets,
and the second term involves the $\cC$-conjugate scalar field
$\tilde\Phi\equiv i\,\sigma_2\,\Phi^*$. In the unitary gauge,
this Yukawa-type Lagrangian takes the simpler form
\bel{eq:y_m}
\cL_Y\, =\, - \left(1+\frac{H}{v}\right)\,\left\{ m_d \,\bar d d + m_u
\,\bar u u + m_\ell \,\bar \ell \ell\right\}\, ,
\ee
with
\bel{eq:f_masses}
 m_f\, = \, y_f\, {v\over\sqrt{2}} \; , \qquad\qquad (f = d, u, \ell)\, .
\ee
Therefore, the EWSB mechanism generates also the masses of the fermion fields.
Since $y_f$ are free parameters, one cannot predict the numerical values of $m_f$. Note, however, that all the Higgs Yukawa couplings are fixed in terms of the measured fermion masses.

\section{Experimental Knowledge on the Higgs Properties}
\label{sec:Higgs-data}

The Higgs interactions have a very characteristic form: they are always proportional to the mass (mass squared) of the coupled fermion (boson), normalized by the vacuum expectation value $v$.
Therefore, the Higgs decay is dominated by tree-level modes with the heaviest kinematically-allowed final states or loop processes involving the top quark.
% All Higgs couplings are determined by $M_H$, $M_W$, $M_Z$, $m_f$ and the vacuum expectation value $v$.
With the measured Higgs mass in Eq.~\eqn{eq:Higgs_Mass}, there is an interesting variety of accesible decay branching fractions; their SM predictions are given in Table~\ref{tab:HiggsBr}.

%%%%%%%%%%%%%%%%%%%%%% Predicted Higgs Branching Fractions %%%%%%%%%%%%%%%%%%%%%%%%%%%%%%
\begin{table}[b]
\centering
\renewcommand{\arraystretch}{1.3} % enlarge line spacing
\renewcommand{\tabcolsep}{1pc} % enlarge column spacing
\begin{tabular}{cc|cc}
\hline
Decay Mode & Br (\%) & Decay Mode & Br (\%)
\\ \hline
$H\to bb$ & $57.5\pm 1.9\phantom{6}$ &
$H\to ZZ^*$ & $2.67\pm 0.11$
\\
$H\to WW^*$ & $21.6\pm 0.9\phantom{6}$ &
$H\to \gamma\gamma$ & $0.228\pm 0.011$
\\
$H\to gg$ & $8.56\pm 0.86$ &
$H\to Z\gamma$ & $0.155\pm 0.014$
\\
$H\to \tau\tau$ & $6.30\pm 0.36$ &
$H\to \mu\mu$ & $0.022\pm 0.001$
\\
$H\to cc$ & $2.90\pm 0.35$ &&
\\\hline
\end{tabular}
\vskip .1cm
\caption{SM predictions for the Higgs decay branching fractions \cite{Heinemeyer:2013tqa,HiggsCombination}.} \label{tab:HiggsBr}
\end{table}
%%%%%%%%%%%%%%%%%%%%%%%%%%%%%%%%%%%%%%%%%%%%%%%%%%%%%%%%%%%%%%%%%%%%%%%

The Higgs boson data are conveniently expressed in terms of the so-called signal strengths, which measure the product of the Higgs production cross section times its decay branching ratio into a given final state, in units of the corresponding SM prediction:
$\mu \equiv \sigma\cdot \mathrm{Br}/(\sigma_{_{\mathrm{SM}}}\cdot \mathrm{Br}_{_{\mathrm{SM}}})$. The SM corresponds to $\mu=1$.
Table~\ref{tab:LHCdata} summarizes the ATLAS %\cite{Aad:2015gba}
and CMS %\cite{CMS:2014ega}
combined measurements \cite{HiggsCombination}, based on the full Run-1 data samples collected at the LHC. These results are in good agreement with the SM.

%%%%%%%%%%%%%%%%%%%%%% Higgs Signal Strengths %%%%%%%%%%%%%%%%%%%%%%%%%%%%%%
\begin{table}[tb]
\centering
\renewcommand{\arraystretch}{1.3} % enlarge line spacing
\renewcommand{\tabcolsep}{1pc} % enlarge column spacing
\begin{tabular}{lccc}
\hline
Decay Mode & ATLAS & CMS & Combined
\\ \hline
$H\to \gamma\gamma$ & $1.15\,{}^{+\, 0.27}_{-\, 0.25}$ & $1.12\,{}^{+\, 0.25}_{-\, 0.23}$
& $1.16\,{}^{+\, 0.20}_{-\, 0.18}$
\\
$H\to ZZ^*$ & $1.51\,{}^{+\, 0.39}_{-\, 0.34}$ & $1.05\,{}^{+\, 0.32}_{-\, 0.27}$
& $1.31\,{}^{+\, 0.27}_{-\, 0.24}$
\\
$H\to WW^*$ & $1.23\,{}^{+\, 0.23}_{-\, 0.21}$ & $0.91\,{}^{+\, 0.24}_{-\, 0.21}$
& $1.11\,{}^{+\, 0.18}_{-\, 0.17}$
\\
$H\to \tau\tau$ & $1.41\,{}^{+\, 0.40}_{-\, 0.35}$ & $0.89\,{}^{+\, 0.31}_{-\, 0.28}$
& $1.12\,{}^{+\, 0.25}_{-\, 0.23}$
\\
$H\to bb$ & $0.62\,{}^{+\, 0.37}_{-\, 0.36}$ & $0.81\,{}^{+\, 0.45}_{-\, 0.42}$
& $0.69\,{}^{+\, 0.29}_{-\, 0.27}$
\\ \hline
Combined & $1.20\,{}^{+\, 0.15}_{-\, 0.14}$ & $0.98\,{}^{+\, 0.14}_{-\, 0.13}$
& $1.09\,{}^{+\, 0.11}_{-\, 0.10}$
\\\hline
\end{tabular}
\vskip .1cm
\caption{Measured Higgs Signal Strengths \cite{HiggsCombination}.} \label{tab:LHCdata}
\end{table}
%%%%%%%%%%%%%%%%%%%%%%%%%%%%%%%%%%%%%%%%%%%%%%%%%%%%%%%%%%%%%%%%%%%%%%%

%%%%%%%%%%%%%%%%%%%%%%% Figure Higgs Production %%%%%%%%%%%%%%%%%%%%%%
\begin{figure}[t]
\centering
\includegraphics[width=12.6cm,clip]{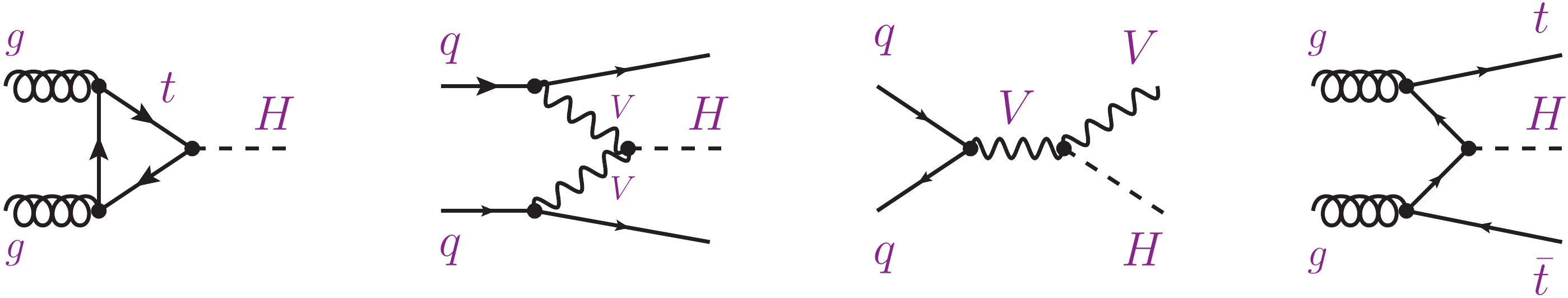}
\caption{Higgs-production mechanisms: $gg\mathrm{F}$, VBF, $V\! H$ and $t\bar t H$.}
\label{fig:HiggsProduction}
\end{figure}
%%%%%%%%%%%%%%%%%%%%%%%%%%%%%%%%%%%%%%%%%%%%%%%%%%%%%%%%%%%%%%%%%%%%%%

The sensitivity to the different Higgs couplings is increased disentangling the different production channels shown in Fig.~\ref{fig:HiggsProduction}: gluon fusion ($gg\mathrm{F}$: $GG\to t\bar t\to H$), vector-boson fusion (VBF: $VV\to H$ ; $V=W,Z$) and associated $VH$ or $t\bar t H$ production. At the LHC, the dominant contribution (86\% at $\sqrt{s}=8$~TeV) comes from the $gg\mathrm{F}$ mechanism, through a triangular quark loop which gives access to the top Yukawa. Owing to the fermion mass dependence of the SM Yukawa couplings, the virtual top loop completely dominates; the bottom contribution is much smaller while the lighter quarks only induce tiny corrections.
The agreement of the measured Higgs production cross section with the SM prediction confirms the existence of a top Yukawa coupling with the expected size. Moreover, it excludes the presence of additional fermionic contributions to $gg\mathrm{F}$ production. A fourth quark generation would increase the cross section by a factor of nine, and much larger enhancements
%($\sim 4 T_R^2/T_F^2$)
would result from exotic fermions in higher colour representations, coupled to the Higgs \cite{Ilisie:2012cc}.

%%%%%%%%%%%%%%%%%%%%%%%%% Figures SignalStrenghts %%%%%%%%%%%%%%%%%%%%%%%%
\begin{figure}[t]
\centering
\begin{minipage}[c]{6cm}\centering
\includegraphics[width=6cm]{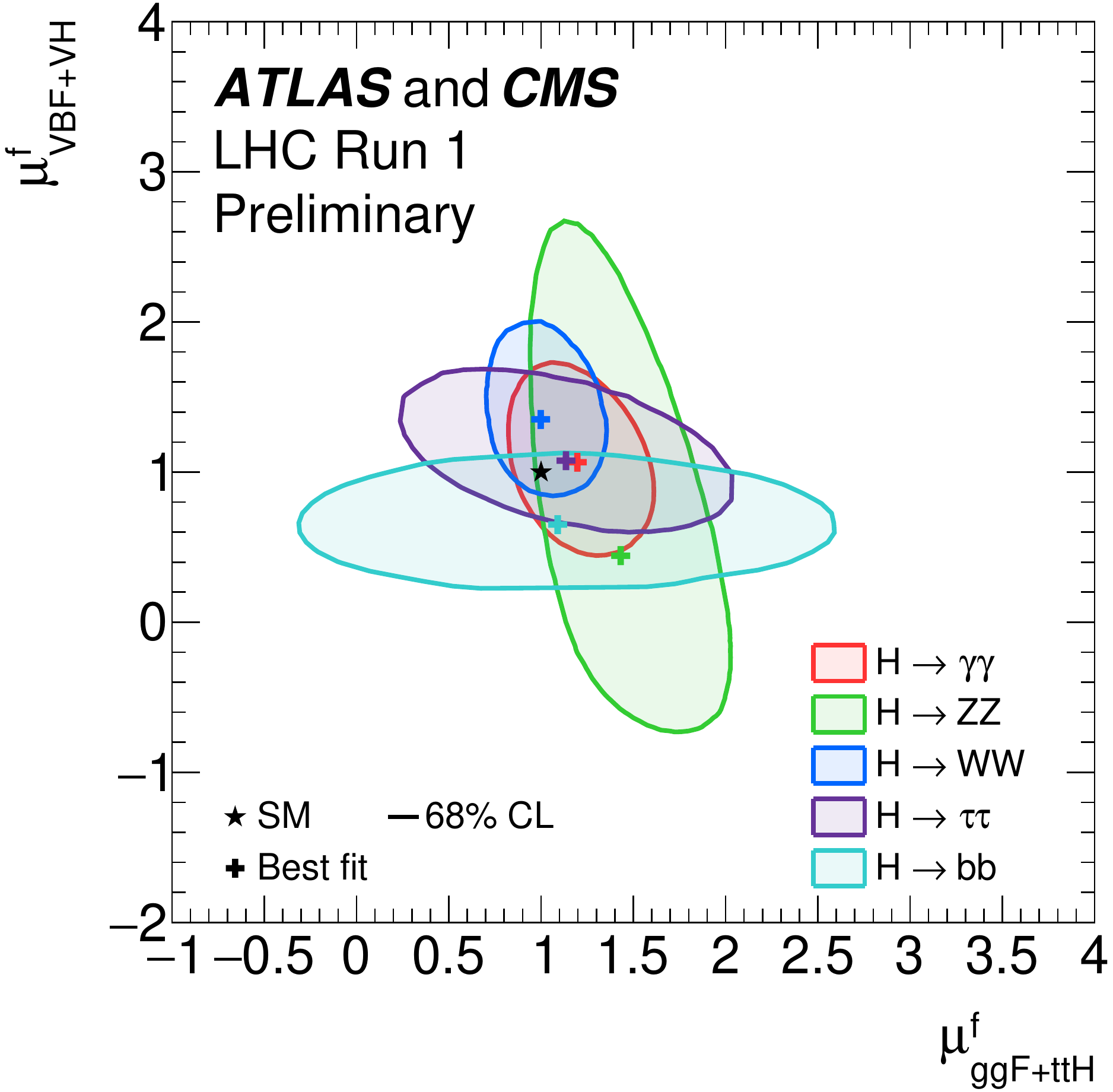}
\caption{Likelihood contours of the production signal strengths for the five measured Higgs decay channels \cite{HiggsCombination}.}
\label{fig:SS_pd}
\end{minipage}
\hfill
\begin{minipage}[c]{6cm}\centering
\includegraphics[width=6cm]{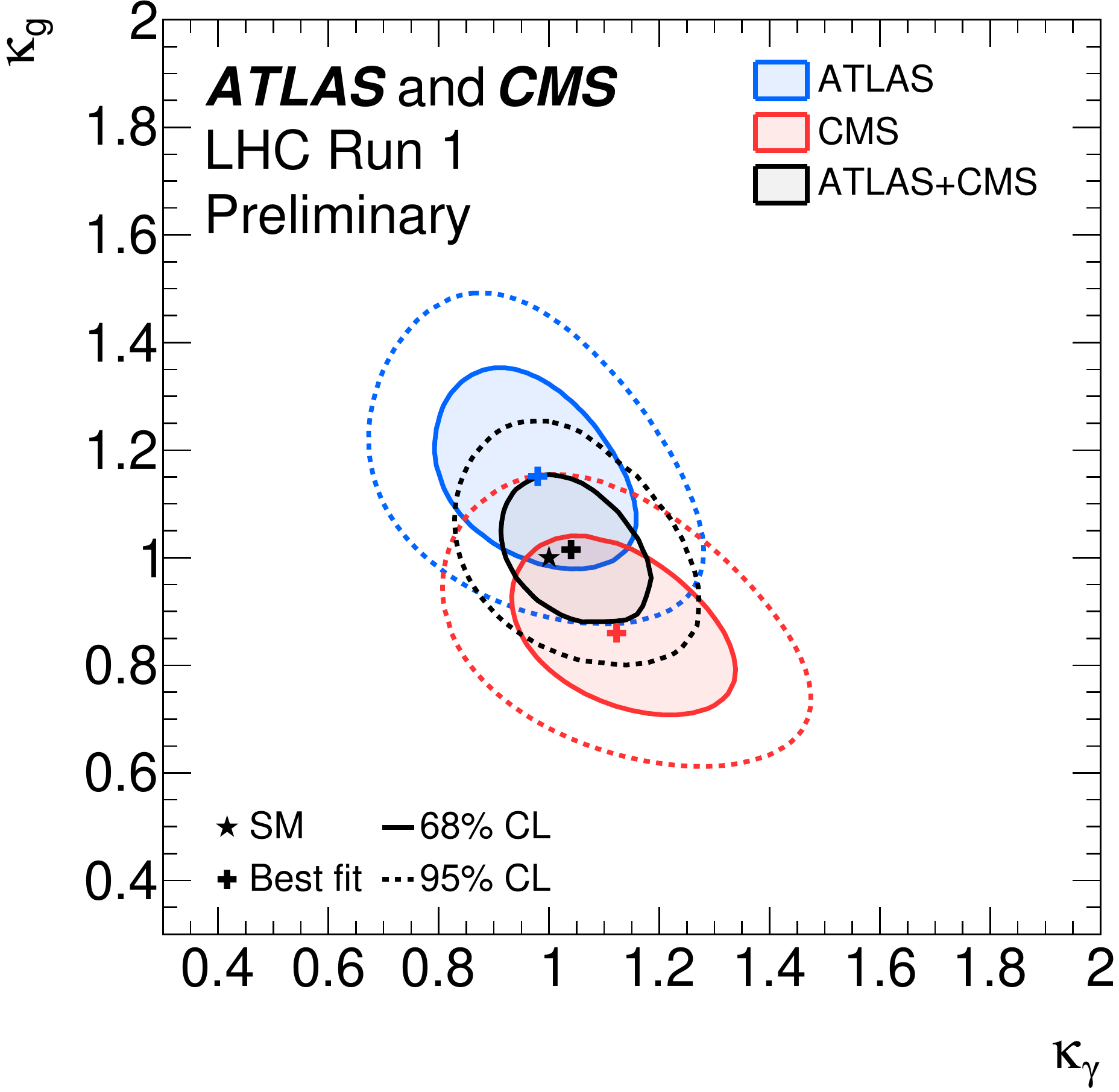}
\caption{Effective $\kappa_g$ and $\kappa_\gamma$ couplings, assuming SM values for all other Higgs interactions \cite{HiggsCombination}.}
\label{fig:kg_kp}
\end{minipage}
\end{figure}
%%%%%%%%%%%%%%%%%%%%%%%%%%%%%%%%%%%%%%%%%%%%%%%%%%%%%%%%%%%%%%%%%%%%%%

Fig.~\ref{fig:SS_pd} shows 68\% CL contours for the five measured signal strengths, separating production mechanisms involving the top Yukawa ($gg\mathrm{F}$ + $t\bar t H$) or the gauge-boson coupling (VBF + $V\! H$). In addition to the dominant $gg\mathrm{F}$ mechanism, there is clear evidence of VBF and $V\! H$ production with a statistical significance of $5.4\,\sigma$ and $3.5\,\sigma$, respectively (assuming SM values for the decay widths) \cite{HiggsCombination}. Actually, $t\bar t H$ production is also seen at the $4.4\,\sigma$ level, but with a too large production signal strength $\mu_{t\bar t H} = 2.3\,{}^{+\, 0.7}_{-\, 0.6}$~\cite{HiggsCombination}.

\begin{figure}[t]
\centering
\includegraphics[width=10cm,clip]{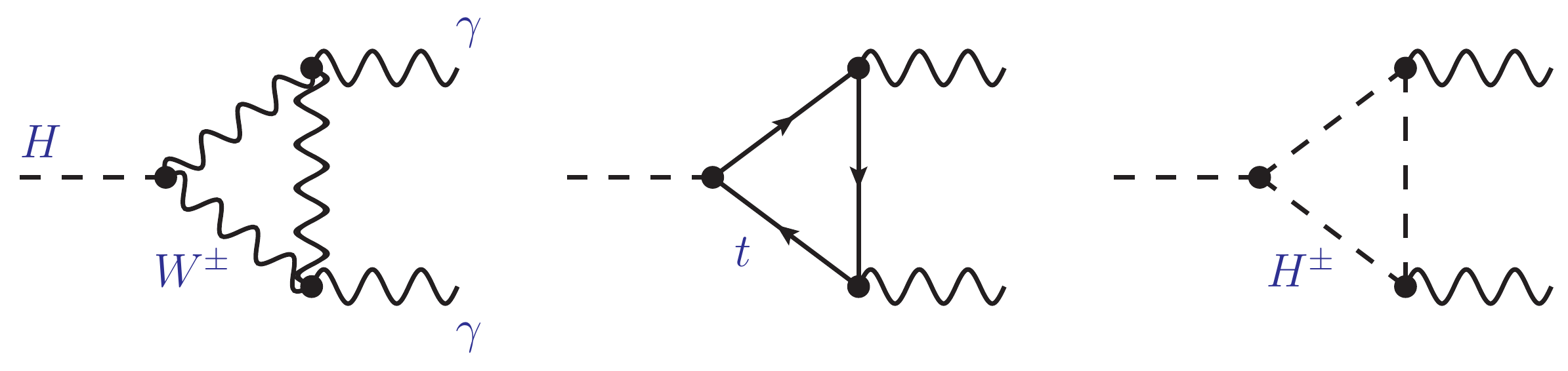}
\caption{One-loop contributions to $H\to \gamma\gamma$. The third diagram shows a possible non-SM contribution from a charged scalar.}
\label{fig:H2gamma}
\end{figure}

The tree-level decays $H\to W^{\pm *} W^{\mp}, Z^* Z$ directly test the electroweak gauge couplings of the Higgs. In addition, we have now strong evidence for the $H$ coupling to $b\bar b$ ($2.6\,\sigma$) and $\tau^+\tau^-$ ($5.5\,\sigma$), through the corresponding fermionic decays~\cite{HiggsCombination}.
The process $H\to\gamma\gamma$ occurs in the SM through intermediate $W^+W^-$ and $t\bar t$
triangular loops, shown in Fig.~\ref{fig:H2gamma}, which interfere destructively; the agreement with the SM prediction confirms the (relative) sign of the top Yukawa.

%%%%%%%%%%%%%%%%%%%%%%%%% Figure H\to 2\gamma %%%%%%%%%%%%%%%%%%%%%%%%
%%%%%%%%%%%%%%%%%%%%%%%%%%%%%%%%%%%%%%%%%%%%%%%%%%%%%%%%%%%%%%%%%%%%%%

The loop amplitudes $H\to\gamma\gamma$ and $H\leftrightarrow gg$ are sensitive to new physics contributions such as the charged-scalar loop in Fig.~\ref{fig:H2gamma}. This is tested in Fig.~\ref{fig:kg_kp} which shows the 68\% and 95\% CL constraints on the effective $H\gamma\gamma$ ($\kappa_\gamma$) and $Hgg$ ($\kappa_g$) couplings, in SM units, assuming that all other Higgs interactions take their SM values. The data are in perfect agreement with the SM point $\kappa_\gamma =\kappa_g = 1$.

\begin{figure}[bt]
\centering
\begin{minipage}[c]{6cm}\centering
\includegraphics[height=5.6cm]{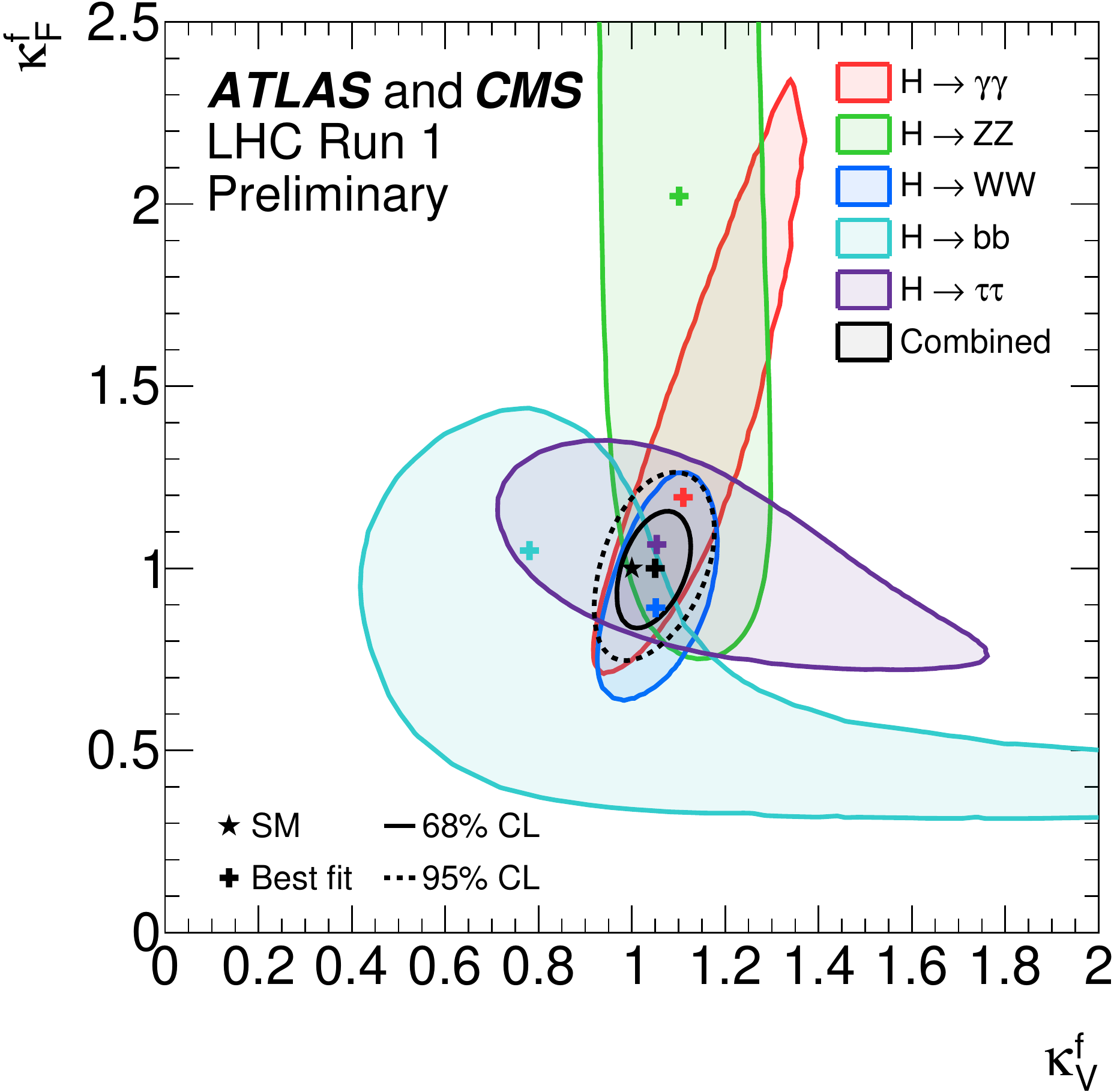}
\caption{Effective bosonic ($\kappa_{_V}$) and fermionic ($\kappa_{_F}$)
Higgs couplings \cite{HiggsCombination}.}
\label{fig:Effcouplings}
\end{minipage}
\hfill
\begin{minipage}[c]{6cm}\centering
\includegraphics[height=5.6cm]{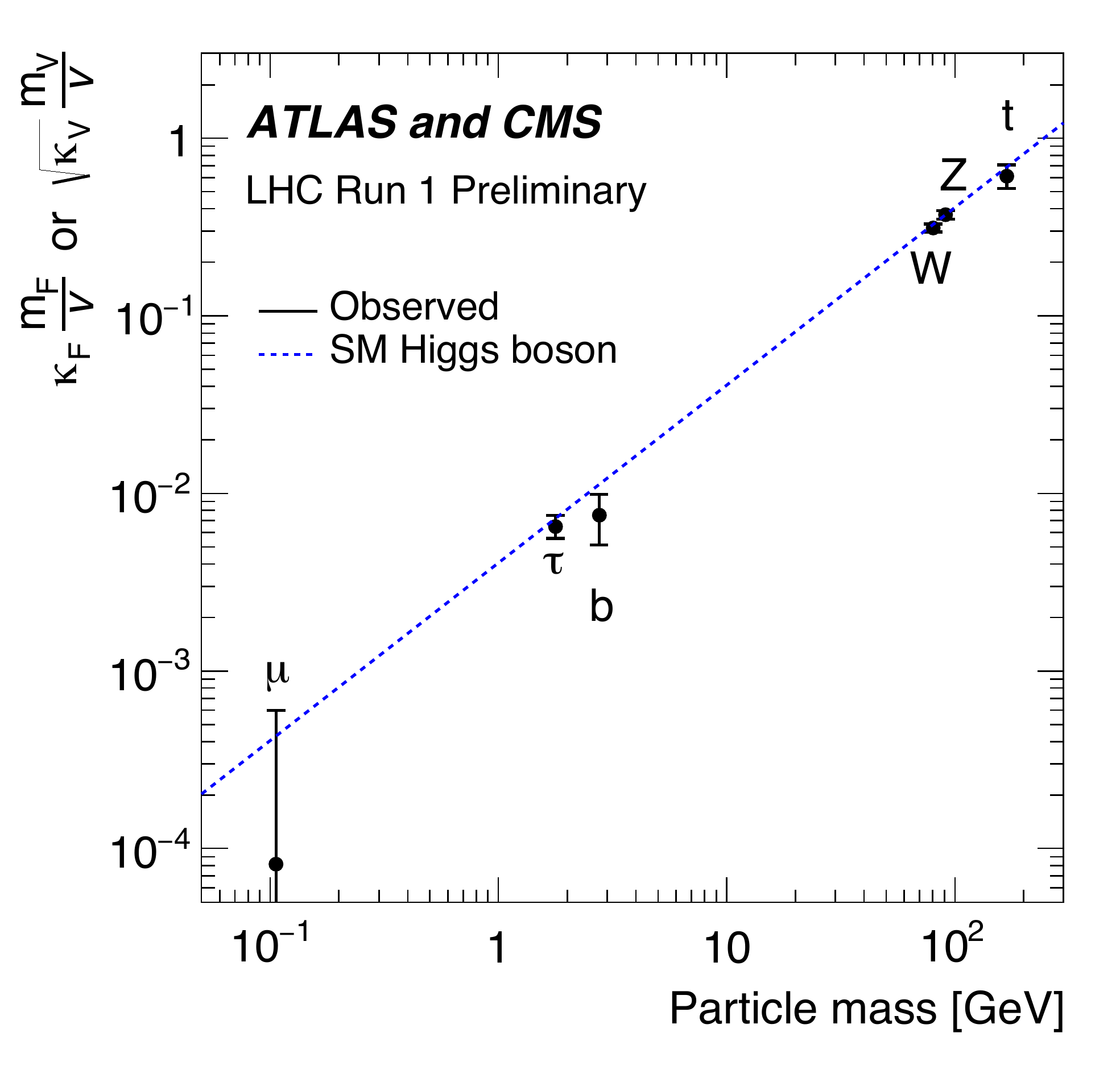}
\caption{Higgs couplings to different particles as function of their masses \cite{HiggsCombination}.}
\label{fig:Hcouplings}
\end{minipage}
\end{figure}

Assuming that there are no new particles in the loops (and the absence of non-SM decay modes), the data can be parametrized in terms of effective Higgs couplings,
$\kappa_n \equiv g_n^{\phantom{\mbox{\tiny SM}}}\!\!\!\! /g_n^{\mbox{\tiny SM}}$.
Taking common vector and fermion coupling modifiers, {\it i.e.},
$\kappa_{_W} = \kappa_{_Z} = \kappa_{_V}$ and $\kappa_t = \kappa_b = \kappa_\tau =\kappa_{_F}$,
Fig.~\ref{fig:Effcouplings} shows the resulting 68\% and 95\% CL constraints for the five measured decay channels. While the tree-level decays are only sensitive to the absolute values of the effective couplings, the $H\to\gamma\gamma$ partial width determines $\kappa_{_F} \kappa_{_V}>0$ (the convention $\kappa_{_V}>0$ has been adopted in the figure).

%%%%%%%%%%%%%%%%%%%%%%%%% Figures Higgs Couplings %%%%%%%%%%%%%%%%%%%%%%%%
%%%%%%%%%%%%%%%%%%%%%%%%%%%%%%%%%%%%%%%%%%%%%%%%%%%%%%%%%%%%%%%%%%%%%%

The mass dependence of the measured Higgs couplings is shown in Fig.~\ref{fig:Hcouplings}, which plots $\kappa_{_F} m_{_F}/v$ (fermions) and $\sqrt{\kappa_{_V}} m_{_V}/v$ (bosons) as function of their masses. The experimental points are in excellent agreement with the SM prediction, indicated by the dashed line.
Moreover, the 95\% CL upper limit $\mathrm{Br}(H\to e^+e^-) < 1.9\times 10^{-3}$ \cite{Khachatryan:2014jba} verifies the strong mass suppression of the electronic coupling. Therefore, the Higgs-like nature of the $H$ boson has been clearly confirmed by the LHC data.

\section{Quantum Loops, Symmetries and the Higgs Mass}
\label{sec:loops}

A fundamental scalar requires some protection mechanism to stabilize its mass. If there is new physics at some heavy scale $\Lambda_{\mathrm{NP}}$, quantum corrections could bring the scalar mass $M_H$ to the new physics scale $\Lambda_{\mathrm{NP}}$:
\bel{eq:MH-corrections}
\delta M_H^2\,\sim\, \frac{g^2}{(4\pi)^2}\, \Lambda_{\mathrm{NP}}^2\,\log{(\Lambda_{\mathrm{NP}}^2/M_H^2)}\, .
\ee
Which symmetry keeps $M_H$ away from $\Lambda_{\mathrm{NP}}$?

Fermion masses are protected by chiral symmetry (invariance under independent phase transformations of the left and right fermion chiralities), while gauge symmetry protects the gauge boson masses. These particles are massless when the symmetry becomes exact.
Therefore, quantum corrections to $m_f$ ($M_W^2$) are necessarily proportional to the fermion ($W$) mass (squared) itself.

This symmetry protection can be also understood through a counting of field components. A massless field with spin $\frac{1}{2}$ has 2 dof while a massive one has 4. Similarly, a massive spin-1 particle has 3 polarizations, but a massless gauge field only contains 2. The massless limit is qualitatively different, making fermion and gauge-boson masses safe against quantum corrections. This is no-longer true for fields without spin structure. A real scalar field has a single component, independently of the value of its mass.

Supersymmetry (SUSY) was originally advocated to protect the Higgs mass. Since it relates boson and fermion fields, combining them in super\-symmetric multiplets, the fermion mass protection is shared with their bosonic partners. However, according to present data this no-longer works `naturally'. SUSY implies a cancellation of fermionic and bosonic quantum corrections to $M_H^2$, which have different signs, but this cancellation is not exact, owing to the necessary presence of SUSY-breaking terms to split the so far undetected sparticle spectrum from the known particle masses. The non-observation of SUSY partners at the LHC indicates that SUSY is badly broken; strong lower bounds on the masses of SUSY particles have been set, surpassing the TeV in many cases. Moreover, the Higgs mass is heavier than what was expected to be naturally accommodated in the minimal SUSY model (MSSM) \cite{Djouadi:2013lra}.

Compositeness is another interesting possibility. Instead of an elementary Higgs, one has a composite bound state made of fermions. The mass of the composite boson state is then governed by the fermion dynamics and symmetries. However, the measured Higgs mass is much lighter than the predictions obtained in the most simplistic scenarios, mostly based on naive extrapolation of QCD physics.

A quite compelling alternative would be a light pseudo-Goldstone Higgs associated with a dynamical breaking of the electroweak symmetry, generated by some underlying strongly-coupled theory \cite{PG-Higgs-1,PG-Higgs-2,PG-Higgs-3,PG-Higgs-4,PG-Higgs-5,PG-Higgs-6,PG-Higgs-7}. One needs a pattern of symmetry breaking $G\to H \subset G$ with at least four broken generators to account for a minimum of 4 Goldstone modes (the 3 SM electroweak Goldstones plus the Higgs). Goldstone bosons are characterized by a Lagrangian shift symmetry:
\be
\varphi_i(x)\;\to\;\varphi_i(x)+c_i
\ee
[see Eqs.~\eqn{eq:doublet} and \eqn{eq:Lphi}]. Therefore, not only the Higgs mass but also the Higgs self-interactions would vanish in this case. These parameters should be generated through quantum effects (or additional symmetry breakings) and would be naturally small. A simple example is provided by the popular $\mathrm{SO(5)}/\mathrm{SO(4)}$ minimal composite Higgs model \cite{Agashe:2004rs,Contino:2006qr}.

The Higgs mass could also be protected by scale symmetry, {\it i.e.}, invariance of the action under transformations of scale:
\be
x^\mu\;\to\; x^\mu/\xi \, , \qquad\qquad\qquad \phi(x)\;\to\; \xi\;\phi(x/\xi)\, .
\ee
In the SM, this symmetry is broken by the quadratic term in the scalar potential which generates the EWSB and the Higgs mass. A scale-invariant SM would only contain massless fields in the Lagrangian. The Higgs-like boson could then arise as a dilaton, the pseudo-Goldstone boson associated with the spontaneous breaking of scale invariance at some scale $f_\varphi\gg v$ \cite{Goldberger:2007zk,Matsuzaki:2012xx,Matsuzaki:2012mk,Bellazzini:2012vz,Chacko:2012vm}.
Although a naive dilaton is basically ruled out by the data, there are other viable implementations of this idea. For instance, one could imagine the existence of
an underlying conformal theory at $\Lambda_{\mathrm{NP}}$; masses should then be generated through quantum effects at lower scales \cite{Coleman:1973jx}.

\section{The Heaviest Mass Scale of the SM}
\label{sec:top}

The top quark is a very sensitive probe of the EWSB, since it is the heaviest fundamental particle within the SM framework. Its large mass,\footnote{%%%%%%%%%%%%%%%%%%
This value is obtained from a kinematical reconstruction of the top decay products and refers to the mass parameter implemented in the Monte Carlo simulations. Although its relation with a well-defined QCD mass is unclear, it is usually identified with the pole of the perturbative quark propagator.
This introduces an additional theoretical uncertainty of the order of 1 GeV  \cite{Hoang:2008xm,Moch:2014tta}.}
%%%%%%%%%%%%%%%%%%%%%%%%%%%%%%%%%%%%%%%%%%%%%%%%%
$m_t = (173.34\pm 0.76)\;\mathrm{GeV}$~\cite{ATLAS:2014wva},
makes the top very different from all other quarks, with a
Yukawa coupling amazingly close to one:
\bel{eq:TopYukawa}
y_t\, = \,\frac{\sqrt{2}}{v}\, m_t\, =\, 2^{3/4} G_F^{1/2} m_t \, =\, 0.995\,\approx\, 1\, .
\ee
For comparison, $y_b\approx 0.025$ and $y_c\approx 0.007 \gg y_{s,d,u}$. One could wonder whether the top quark is really a genuine SM particle. If some (non-perturbative) strong dynamics is responsible for the EWSB, the top should obviously be directly linked to it.

Up to now, top quarks have only been detected through their decay mode $t\to W^+ b$, because
the top couplings to the lighter quark generations are very small. The measured single-top production cross section implies $|V_{tb}| > 0.92 \; (95\%\,\mathrm{CL})$ \cite{Aaltonen:2015cra,Khachatryan:2014iya}.

%%%%%%%%%%%%%%%%%%%%%% Figure EW fit %%%%%%%%%%%%%%%%%%%%%%%%%%%%%%
\begin{figure}[t]
\centering
%\mbox{}\hskip -.05cm
\includegraphics[width=8.5cm,clip]{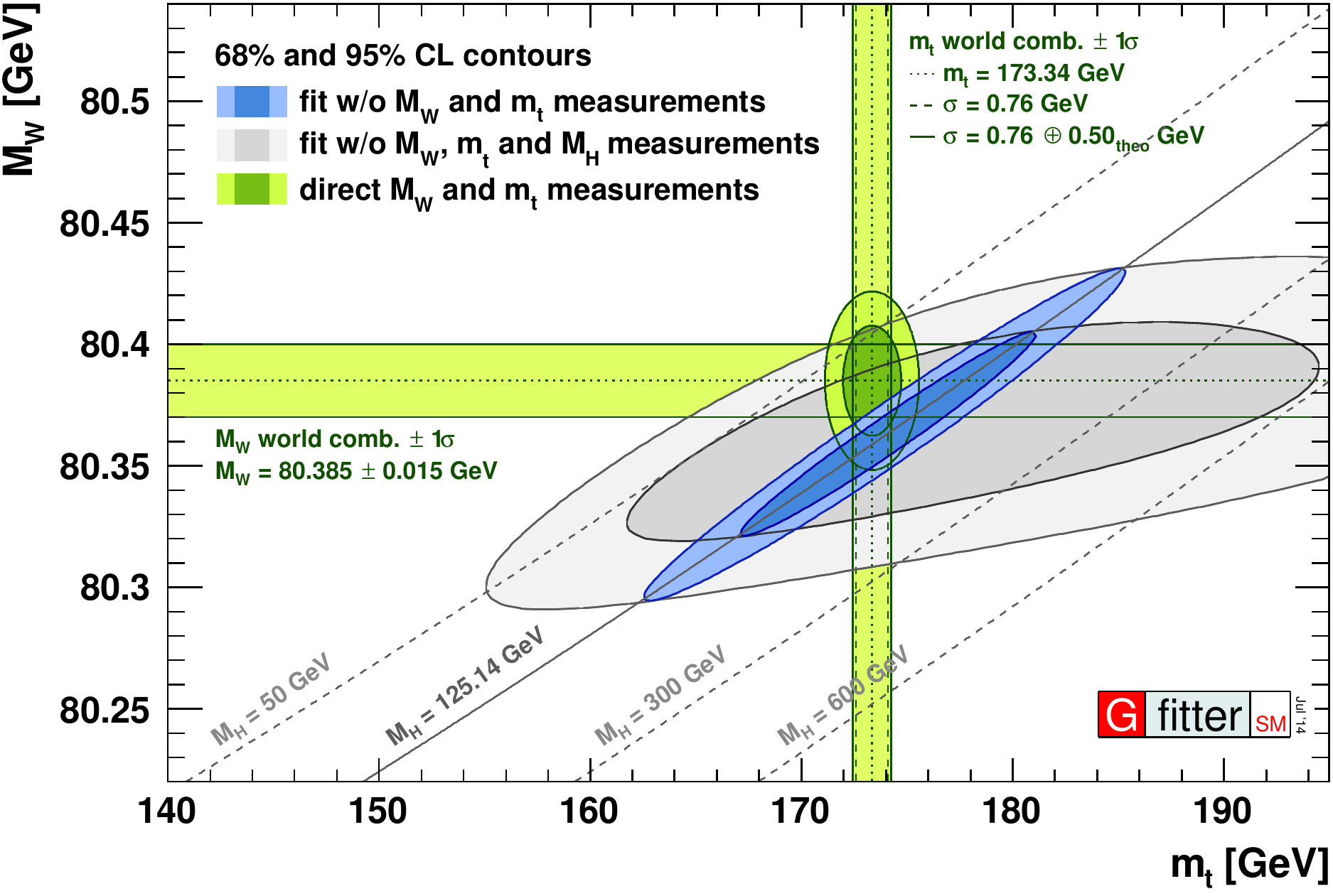}
\caption{SM electroweak fit in the $m_t$--$M_W$ plane,
%(excluding the direct measurements of $m_t$ and $M_W$),
with (blue) and without (gray) $M_H$, compared with the direct measurements of the top and $W$ masses (green) \cite{Baak:2014ora}.}
\label{fig:EWfit_mt}
\end{figure}
%%%%%%%%%%%%%%%%%%%%%%%%%%%%%%%%%%%%%%%%%%%%%%%%%%%%%%%%%%%%%%%%%%%

Virtual top contributions dominate the electroweak quantum corrections to many relevant quantities, such as the $W^\pm$ and $Z$ propagators \cite{Veltman:1977kh} or the $Zb\bar b$ vertex \cite{Bernabeu:1987me,Bernabeu:1990ws}. These effects increase quadratically with the top mass, while virtual Higgs contributions grow logarithmically with $M_H$ in the gauge self-energies and are negligible in $Zb\bar b$. This provides a quite good sensitivity to $m_t$ and $M_H$ through precision electroweak data. As shown in Fig.~\ref{fig:EWfit_mt}, the direct measurements of the Higgs, top and $W^\pm$ masses are in beautiful agreement with the values extracted indirectly from global electroweak fits. This constitutes a very significant test of the SM at the quantum level.

Quantum corrections to $M_H^2$ are also dominated by contributions from top loops, which grow logarithmically with the renormalization scale $\mu$:
\bel{eq:MH_QC}
\frac{M_H^2}{2 v^2} \,\approx\, \lambda(\mu) + \frac{2 y_t^2}{(4\pi)^2}\left[\lambda + 3 (y_t^2-\lambda)\,\log{(\mu/m_t)}\right] .\;
\ee
As expected, $M_H$ is brought close to the heaviest SM mass $m_t= y_t\, v/\sqrt{2}$.
Since the physical value of
$M_H$ is fixed, the tree-level contribution $2 v^2\lambda(\mu)$ decreases with increasing $\mu$. %
%%%%%%%%%%%%%%%%%%%%%%% Figure EW fit %%%%%%%%%%%%%%%%%%%%%
\begin{figure}[t]
\centering
\includegraphics[width=7.55cm,clip]{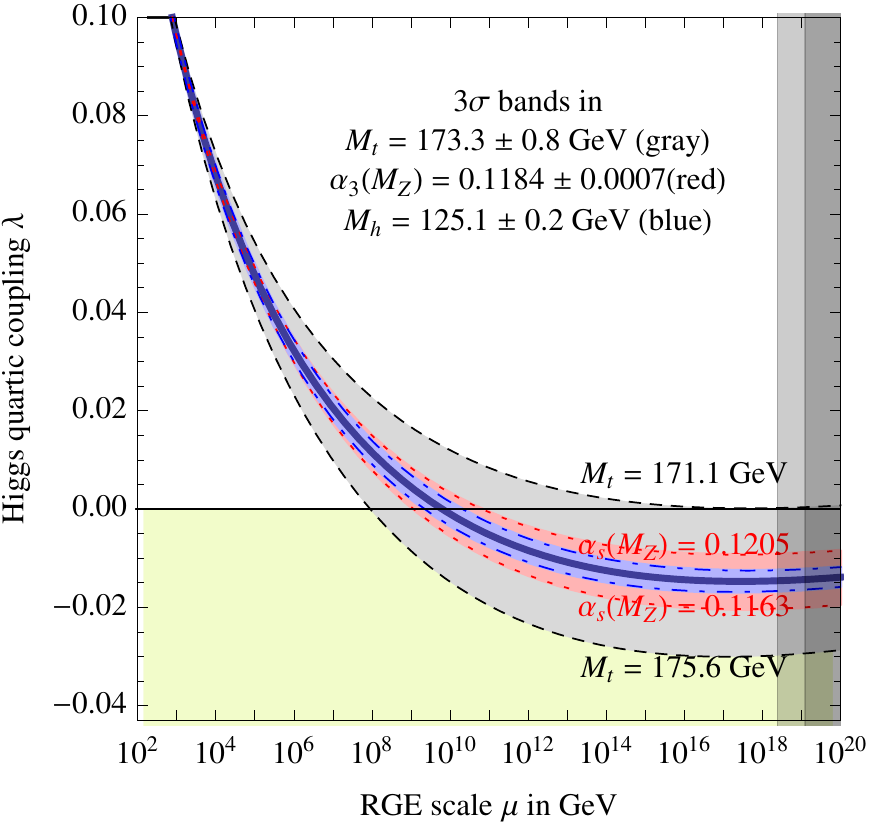}
\caption{Evolution of $\lambda(\mu)$ with the renormalization scale  \cite{Buttazzo:2013uya}.}
\label{fig:lambda}
\end{figure}
%%%%%%%%%%%%%%%%%%%%%%%%%%%%%%%%%%%%%%%%%%%%%%%%%%%%%%%%%%%%%%%%%%%%
%
Fig.~\ref{fig:lambda} shows the evolution of $\lambda(\mu)$ up to the Planck scale ($M_{\mathrm{Pl}} = 1.2\times 10^{19}$~GeV), at the next-to-next-to-leading order, varying $m_t$, $\alpha_s(M_Z^2)$ and $M_H$ by $\pm 3\sigma$ \cite{Buttazzo:2013uya}. The quartic coupling remains weak in the entire energy domain below $M_{\mathrm{Pl}}$ and crosses $\lambda=0$ at very high energies, around $10^{10}$~GeV. The values of $M_H$ and $m_t$ are very close to those needed for absolute stability of the potential ($\lambda >0$) up to  $M_{\mathrm{Pl}}$, which would require $M_H > (129.6\pm 1.5)$~GeV \cite{Buttazzo:2013uya}
($\pm 5.6$~GeV with more conservative errors on $m_t$ \cite{Alekhin:2012py}). Even if $\lambda$ becomes slightly negative at very high energies, the resulting potential instability leads to an electroweak vacuum lifetime much larger than any relevant astrophysical or cosmological scale. Thus, the measured Higgs and top masses result in a metastable vacuum \cite{Buttazzo:2013uya} and the SM could be valid up to $M_{\mathrm{Pl}}$. The possibility of some new-physics threshold at scales $\Lambda\sim M_{\mathrm{Pl}}$, leading to the matching condition
 $\lambda(\Lambda) = 0$, is obviously intriguing.

\section{Scalar-Singlet Extension of the SM}
\label{sec:xSM}

The relation $M_W = M_Z \cos{\theta_W}$ is a very successful prediction of the SM Higgs mechanism, which originates in the doublet structure of the SM scalar field. An extended scalar sector with several fields $\Phi_i$ belonging to different $SU(2)_L\otimes U(1)_Y$ representations $(T_i, Y_i)$ would lead in general to a different relation between the gauge-boson masses. At tree level, one easily gets the result
\bel{eq:RhoParameter}
\rho\, \equiv\, \frac{M_W^2}{M_Z^2 \cos^2{\!\theta_W}}\, =\, \frac{\sum_i v_i^2\,\left[ T_i (T_i + 1) - Y_i^2\right]}{2\,\sum_i v_i^2\, Y_i^2}\, ,
\ee
with $v_i/\sqrt{2}$ the vacuum expectation value of the neutral field component of the $\Phi_i$ multiplet. In the SM, with a single scalar doublet ($T = \frac{1}{2}$) of hypercharge $Y=\frac{1}{2}$, one gets $\rho = 1$. The same prediction would obviously be obtained adding an arbitrary number of doublets with $Y=\frac{1}{2}$ and singlet fields ($T_i=Y_i=0$). Scalar multiplets in higher $SU(2)_L$ representations would result in $\rho \not= 1$, unless their hypercharges are conveniently tuned to get the desired result. Therefore, doublets and singlets are the favoured candidates for building alternative models of perturbative EWSB.

The simplest extension of the SM scalar sector is provided by the addition of a real scalar field $S$, singlet under the SM gauge group. The scalar potential takes the form:
\beqn\label{eq:V_xSM}
V(\Phi,S)& =& \lambda\,\left(|\Phi|^2-\frac{v^2}{2}\right)^2
+ \left( a_\Phi\, S + b_\Phi\, S^2\right) \left(|\Phi|^2-\frac{v^2}{2}\right)
\no\\ & + &
\mu_S^2\, S^2 + a_S\, S^3 + \lambda_S\, S^4\, .
\eeqn
A possible linear term in $S$ has been eliminated through a redefinition of the singlet field.
With this parametrization, the minimum of the scalar potential is obtained at
$\langle 0 | S | 0\rangle = 0$ and $\langle 0 | \phi^{(0)}| 0\rangle = v/\sqrt{2}$, provided that $4\lambda\,\mu_S^2 > a_\Phi^2 \ge 0$. Requiring a positive growing of the potential at large field values implies the conditions $\lambda, \lambda_S, b_\Phi > 0$.

The physical spectrum of the model contains two neutral scalars.
The potential $V(\Phi,S)$ mixes the singlet scalar field $S$ with the
neutral component of the scalar doublet $\phi^{(0)} = \frac{1}{\sqrt{2}}\, (v + \hat\phi)$.
Diagonalizing the terms quadratic in the fields, one easily finds the mass eigenstates:
\bel{eq:xSM_Mixing}
\left(\! \ba h\\[3pt] H\ea\!\right)\, =\, \left[ \bat \cos{\alpha} & \sin{\alpha}\\[3pt]
 -\sin{\alpha} & \cos{\alpha}\ea\right] \left( \!\ba \hat\phi\\[3pt] S\ea\!\right)\, ,
\ee
with the mixing angle given by
\bel{eq:xSM_MixingA}
\tan{2\alpha}\, =\, \frac{a_\Phi v}{v^2\lambda - \mu_S^2}\, .
\ee
We adopt the convention $M_h < M_H$ and $-\frac{\pi}{2}\le \alpha\le\frac{\pi}{2}$.
The masses of the two neutral scalars are then:
\bel{eq:xSM_Masses}
M_h^2\, =\, \Sigma -\Delta \qquad < \qquad
M_H^2\, =\, \Sigma +\Delta
\ee
where
\bel{eq:xSM_Masses_2}
\Sigma\, =\, v^2\lambda + \mu_S^2 \, ,
\qquad\qquad\quad
\Delta\, =\, \sqrt{ (v^2\lambda - \mu_S^2)^2 + a_\phi^2 v^2}\, .
\ee

The field $S$ does not couple to fermions and gauge bosons because it is a singlet under $SU(2)_L\otimes U(1)_Y$ transformations. Therefore, the physical scalars $h$ and $H$ only couple to those particles through their doublet component $\hat\phi$, which results in a universal reduction of all their couplings with respect to the SM Higgs:
\beqn\label{xSM_Couplings}
\kappa^h_V\,\equiv\, g_{hVV}^{\phantom{\mbox{\tiny SM}}}/g_{HVV}^{\mbox{\tiny SM}}\, =\,\cos{\alpha}\, , &\qquad\quad &
\kappa^h_f\,\equiv\, y_{hff}^{\phantom{\mbox{\tiny SM}}}/y_{Hff}^{\mbox{\tiny SM}}\, =\,\cos{\alpha}\, ,
\no\\
\kappa^H_V\,\equiv\, g_{HVV}^{\phantom{\mbox{\tiny SM}}}/g_{HVV}^{\mbox{\tiny SM}}\, =\,\sin{\alpha}\, , &&
\kappa^H_f\,\equiv\, y_{Hff}^{\phantom{\mbox{\tiny SM}}}/y_{Hff}^{\mbox{\tiny SM}}\, =\,\sin{\alpha}\, .\quad
\eeqn
The lighter scalar has the same decay branching ratios as the SM Higgs,
$\mathrm{Br}(h\to f) = \mathrm{Br}(h\to f)_{_{\mathrm{SM}}}$, while its total decay width is reduced by a factor $\Gamma_h^{\phantom{\mbox{\tiny SM}}}/ \Gamma_H^{\mbox{\tiny SM}} = \cos^2{\!\alpha}$. For the heavier scalar $H$, one must take also into account the decay $H\to hh$, which is allowed for $M_H > 2 M_h$. Thus, one gets the signal strengths:
\bel{eq:xSM_SS}
\mu_h\, =\, \cos^2{\!\alpha}\, ,
\qquad\qquad
\mu_{H\to VV, f\bar f}\, =\, \sin^2{\!\alpha}\,\left[ 1 -\mathrm{Br}(H\to hh)\right]\, .
\ee
%

%%%%%%%%%%%%%%%%%%%%%%% Figure xSM %%%%%%%%%%%%%%%%%%%%%
\begin{figure}[t]
\centering
\includegraphics[width=12.5cm,clip]{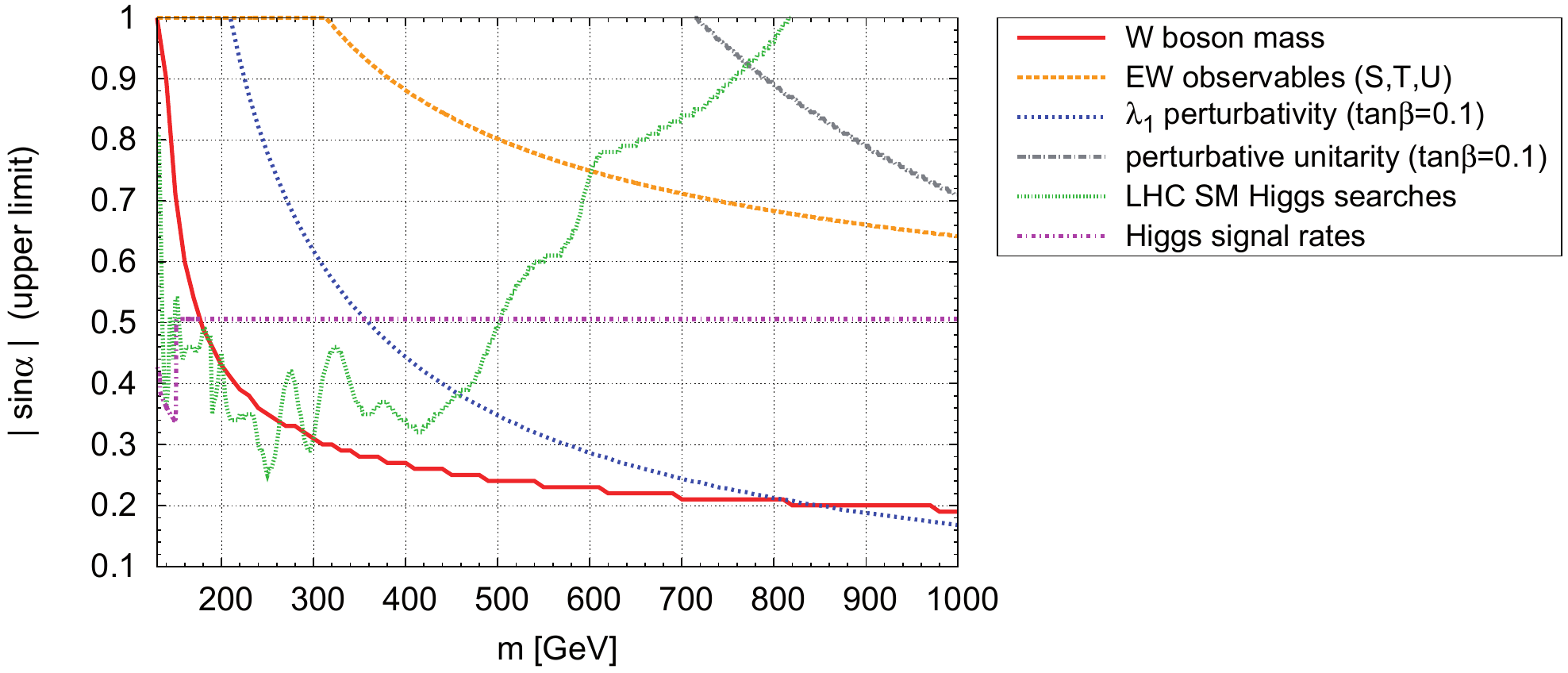}
\caption{95\% CL upper limits on the neutral mixing factor $|\sin{\alpha}|$, as a function of the heavier scalar mass $M_H$ \cite{Robens:2015gla}.}
\label{fig:xSM}
\end{figure}
%%%%%%%%%%%%%%%%%%%%%%%%%%%%%%%%%%%%%%%%%%%%%%%%%%%%%%%%%%%%%%%%%%%%

Fig.~\ref{fig:xSM} shows the present phenomenological constraints on the mixing factor $\sin{\alpha}$ \cite{Robens:2015gla}, assuming that the lighter scalar state corresponds to the discovered Higgs boson, {\it i.e.}, $M_h = 125$~GeV. The measured Higgs signal strengths imply a direct lower bound on $\cos^2{\!\alpha}$ and a corresponding upper limit on $|\sin{\alpha}|$ which is indicated by the magenta horizontal line (a slightly better limit is obtained with the updated values in Table~\ref{tab:LHCdata}).
The green curve displays the bounds from direct heavy Higgs searches.
The figure shows also some theoretical constraints obtained with the requirements of perturbative unitarity (grey) and a perturbative $\lambda$ coupling (blue), taking
$\tan{\beta} \equiv 4 v \lambda_S/a_S = 0.1$. The yellow curve shows the constraints from global electroweak precision fits to the gauge-boson self-energies, parametrized through the so-called oblique parameters $S$, $T$ and $U$.

\begin{figure}[t]
\centering
\includegraphics[width=5cm,clip]{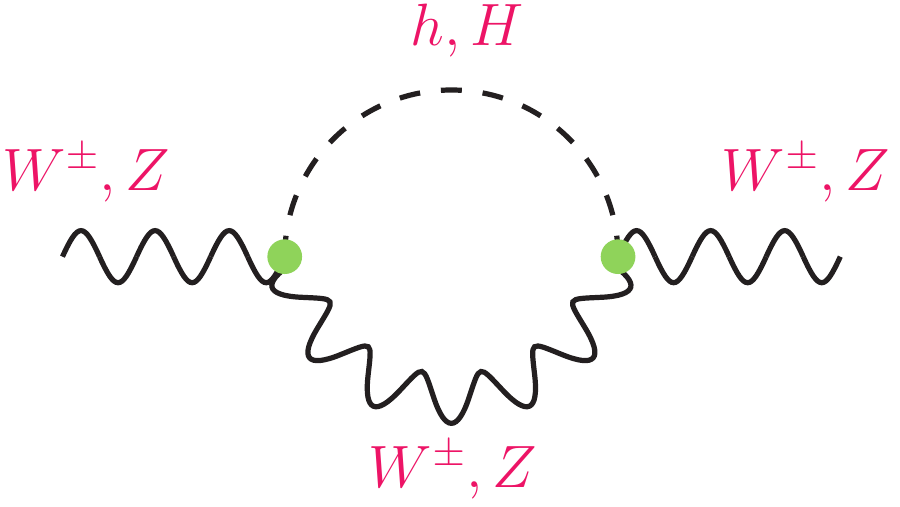}
\caption{Scalar loop contributions to the gauge boson self-energies.}
\label{fig:GaugeSE}
\end{figure}

The most stringent limit (red line) \cite{Lopez-Val:2014jva} comes from the precise experimental measurement of $M_W$. The neutral scalars contribute to the $Z$ and $W^\pm$ self-energies through the loop diagrams shown in Fig.~\ref{fig:GaugeSE}, generating a quantum correction $\Delta r$ to the relation  ($M_Z$ and $\alpha$ are SM input parameters)
\bel{eq:Dr}
M_W^2\,\left( 1 -\frac{M_W^2}{M_Z^2}\right)\, =\,
\frac{\pi\alpha}{\sqrt{2}\, G_F}\; \left( 1 + \Delta r\right)\, .
\ee
Compared with the SM, the additional loop contributions scale with $\sin^2{\!\alpha}$,
\bel{eq:xSM-scaling}
%\delta(\Delta r)
\Delta r - \Delta r^{\mbox{\tiny SM}}
\; =\; \underbrace{\Delta r^H}_{\sin^2{\alpha}}\, +\,
\underbrace{\Delta r^h - \Delta r^{\mbox{\tiny SM}}}_{\cos^2{\!\alpha}\, -\, 1}\;\;
\propto\; \sin^2{\!\alpha}\, ,
\ee
which allows one to set a tight constraint on the scalar mixing angle.
Thus, a very heavy neutral scalar above 700~GeV should necessarily have quite suppressed couplings to the fermions and gauge bosons.

%%%%%%%%%%%%%%%%%%%%%%% Figure xSM %%%%%%%%%%%%%%%%%%%%%
%%%%%%%%%%%%%%%%%%%%%%%%%%%%%%%%%%%%%%%%%%%%%%%%%%%%%%%%%%%%%%%%%%%%

A particularly interesting possibility is the unmixed case $\alpha=0$, which occurs when $a_\Phi=0$. The absence of mixing can be guaranteed at the quantum level imposing the Lagrangian to be symmetric under the discrete ${\cal Z}_2$ transformation $S\to -S$, which excludes terms linear or cubic in the singlet field, {\it i.e.}, $a_\phi = a_S = 0$. The singlet scalar becomes then an absolutely stable dark matter (DM) candidate, which only communicates with the rest of the world through its coupling with the scalar doublet $b_\Phi\, S^2 |\Phi|^2$
(Higgs portal) \cite{Silveira:1985rk}. This provides the simplest example of an ultraviolet-complete theory containing a weakly-interacting massive particle (WIMP) as a viable candidate for DM. This model satisfies all present experimental constraints (scattering of $S$ on nucleons through Higgs exchange, annihilation into SM particles through $SS\to HH$, invisible Higgs decay width) and is able to reproduce the observed DM relic density with natural values of $|b_\Phi|\lesssim 1$ and $M_S$ below a few TeV, predicting at the same time a cross section for scattering on nucleons that is not far from the current direct detection limit \cite{Cline:2013gha}.

\section{Multi-Higgs-Doublet Models}
\label{sec:mHDM}

Let us consider an extended scalar sector involving $N$ $SU(2)_L$ doublets $\phi_a$ with hypercharge $Y=\frac{1}{2}$. It contains $4 N$ real scalars; 3 of them are needed as electroweak Goldstones, giving rise to the longitudinal polarizations of the gauge bosons, and $4N-3$ Higgses remain in the physical spectrum: $(N-1)$ positively charged particles with their corresponding negatively charged antiparticles and $2N-1$ neutral bosons.
The rich variety of fields provides a broad range of dynamical possibilities with very
interesting phenomenological implications.

The neutral components of the scalar doublets acquire vacuum expectation values, which in general could be complex:
$\langle 0|\,\phi_a^T(x)|0\rangle = \frac{1}{\sqrt{2}}\, (0\, , v_a\,\e^{i\theta_a})$.
Without loss of generality, we can enforce $\theta_1\! =\! 0$
through an appropriate $U(1)_Y$ transformation.
It is convenient to perform a global $SU(N)$ transformation in the space of scalar fields,
so that only one doublet acquires non-zero vacuum expectation value
$v\equiv\left(v_1^2 + \cdots + v_N^2\right)^{1/2}$. This defines the
so-called Higgs basis, with the doublets parametrized as
\bel{eq:Higgs_Basis}
\Phi_1 \, = \,\left[ \ba G^+\\ \frac{1}{\sqrt{2}}\left( v + S_1 + i\, G^0\right)\ea \right]\, ,
\qquad\qquad
\Phi_{a>1} \, = \,\left[ \ba H_a^+\\ \frac{1}{\sqrt{2}}\left( S_a + i P_a\right)\ea \right] \, .
\ee
In this basis the Goldstone fields $G^\pm(x)$ and $G^0(x)$ get isolated as components of $\Phi_1$, which is the only doublet responsible for the EWSB and plays the same role as the SM Higgs. The physical charged (neutral) mass eigenstates are linear combinations of the scalar fields $H_a^\pm$ ($S_a$ and $P_a$), determined by the scalar potential.

The scalar field $S_1$ couples to the gauge bosons in exactly the same way as the SM Higgs, {\it i.e.}, its gauge couplings are given by Eq.~\eqn{eq:L_H} with $H$ replaced by $S_1$. Writing $S_1$ in terms of neutral mass eigenstates $\varphi_i^{0}$, $S_1 = \sum_i {\cal R}_{i1} \varphi_i^{0}$, one gets
$|g_{\varphi_i^{0}VV}/g^{\mbox{\tiny SM}}_{HVV}|= |{\cal R}_{i1}|\le 1$.
Moreover, the orthogonality of the field transformation ${\cal R}_{ij}$ implies
\bel{eq:orthogonality}
\sum_{i=1}^{2N-1} g_{\varphi_i^{0}VV}^2 \, =\, \left(g^{\mbox{\tiny SM}}_{HVV}\right)^2\, .
\ee
Thus, the gauge coupling of the SM Higgs is shared among the fields $\varphi_i^{0}$.

All scalar doublets can couple to the fermion fields.
The most general Yukawa Lagrangian with $N$ scalar doublets takes the form
\bel{eq:GenYukawa}
\cL_Y \, =\,\mbox{} -\sum_{a=1}^N\;\left\{ \bar Q_L'\left(
\cY^{(a)'}_d\phi_a\, d_R'
+ \cY^{(a)'}_u\tilde\phi_a\, u_R' \right)
%\right.\CR\!\!\! && \hskip .9cm\Bigl.\mbox{}
+ \bar L_L'\, \cY^{(a)'}_\ell\phi_a\, \ell_R'\right\} + \mathrm{h.c.}\, ,\quad
\ee
where
%$\phi_a(x)$ are the %%% hypercharge
% $Y=\frac{1}{2}$ scalar doublets,
$\tilde\phi_a(x) \equiv i \tau_2\,{\phi_a^*}$,
% are the charge-conjugated scalar fields,
$Q_L'$ and $L_L'$ denote the left-handed quark and lepton doublets and $d'_R$,
$u'_R$ and $\ell'_R$ are the corresponding right-handed fermion singlets.
For simplicity, we do not consider right-handed neutrinos, although they
could be easily incorporated.
All fermionic fields are written as $N_G$-dimensional flavour vectors, with $N_G$ the number of fermion families,
{\it i.e.}, $d_R' = (d_R', s_R', b_R',\cdots )^T$ and similarly for $u'_R$, $\ell_R'$, $Q_L'$ and $L_L'$.
The couplings $\cY^{(a)'}_f$ ($f=d,u,\ell$) are then
$N_G\times N_G$ complex matrices in flavour space.

The physics of these Yukawa interactions becomes more transparent in the Higgs basis:
\bel{eq:LY_Hbasis}
\sum_{a=1}^N\, \cY^{(a)'}_{d,\ell}\,\phi_a\, =\,\sum_{a=1}^N\, Y^{(a)'}_{d,\ell}\,\Phi_a\, ,
\qquad\qquad
\sum_{a=1}^N\, \cY^{(a)'}_{u}\,\tilde\phi_a\, =\,\sum_{a=1}^N\, Y^{(a)'}_{u}\,\tilde\Phi_a\, .
\ee
The fermion masses originate from the $\Phi_1$ couplings, because $\Phi_1$ is the only scalar field acquiring a vacuum expectation value:
\bel{eq:FermMass}
M'_f\, =\, Y^{(1)'}_f\, \frac{v}{\sqrt{2}}\, .
\ee
The diagonalization of the fermion mass matrices $M'_f$ ($f=d,u,\ell$) defines the fermion mass eigenstates $u_i$, $d_i$, $\ell_i$ ($i = 1,\cdots, N_G$) with diagonal mass matrices $M_f$.
In the SM, with only one scalar doublet, this automatically diagonalizes the Higgs interactions which are flavour conserving. In the fermion mass-eigenstate basis, the $Z$ couplings are also diagonal (GIM mechanism \cite{Glashow:1970gm}) and the only source of flavour-changing transitions is the Cabibbo-Kobayashi-Maskawa (CKM) \cite{Cabibbo:1963yz,Kobayashi:1973fv} quark mixing matrix appearing in the $W^\pm$ interactions.
The extraordinary phenomenological success of the SM description of flavour is deeply rooted in the unitarity structure of the CKM matrix and the absence of any flavour-changing vertices in the interactions of the neutral fields \cite{Pich:2011nh}.

When more scalar doublets are present ($N> 1$), there are several Yukawa matrices $Y^{(a)'}_f$ coupled to the same type of right-handed fermions $f_R$. In general, one cannot diagonalize simultaneously all these matrices. Therefore, in the fermion mass-eigenstate basis
the matrices $Y^{(a)}_f$ with $a\not=1$ remain non-diagonal giving rise to dangerous flavour-changing transitions mediated by the neutral scalars.
The appearance of flavour-changing neutral-current (FCNC) interactions represents a major
shortcoming of the model. Since FCNC phenomena are experimentally tightly constrained, one needs to implement ad-hoc dynamical restrictions to guarantee the suppression of the FCNC couplings at the required level. Unless the Yukawa couplings are very small or the scalar bosons very heavy, a very specific flavour structure is required by the data.

\subsection{Flavour Alignment}

The unwanted non-diagonal neutral couplings can be eliminated requiring the alignment
in flavour space of the Yukawa matrices \cite{Pich:2009sp,Pich:2010ic}:
\beqn\label{eq:alignment}
Y^{(a)'}_{d,\ell}  & =&  \varsigma_{d,\ell}^{(a)}\, Y^{(1)'}_{d,\ell}\; =\; \frac{\sqrt{2}}{v}\; \varsigma_{d,\ell}^{(a)}\, M'_{d,\ell}\, ,
\no\\
Y^{(a)'}_{u}  & =&  \varsigma_{u}^{(a)*}\, Y^{(1)'}_{u}\; =\; \frac{\sqrt{2}}{v}\; \varsigma_{u}^{(a)*}\, M'_{u}\, ,
\eeqn
where $\varsigma_{f}^{(a)}$ are complex proportionality parameters
($\varsigma_{f}^{(1)}=1$).
All Yukawa matrices coupling to a given type of right-handed fermions are assumed to be proportional to each other and can, therefore, be diagonalized simultaneously.
In terms of fermion mass eigenstates, $\cL_Y$ takes then the form:
\beqn\label{eq:alignedY}
\cL_Y & =&
-\frac{\sqrt{2}}{v}\;\sum_{a=2}^N\, H^+_a \left\{
\varsigma_{d}^{(a)} \,\bar u_L V M_{d} d_R
\,-\,\varsigma_{u}^{(a)} \,\bar u_R M_{u}^\dagger V d_L \,
+ \,\varsigma_{\ell}^{(a)} \,\bar\nu_L  M_\ell\,\ell_R\right\}
\no\\ &&
- \sum_f\; \bar f_L M_f f_R\; \left\{ 1 + \frac{1}{v}\; S_1\right\}
\, -\, \frac{1}{v}\; \sum_{a=2}^N\, \left[ S_a - i P_a\right]\,
\varsigma_{u}^{(a)*} \,\bar u_L M_{u} u_R
\no\\ &&
-\frac{1}{v}\; \sum_{a=2}^N\, \left[ S_a + i P_a\right]\left\{
\varsigma_{d}^{(a)} \,\bar d_L M_{d} d_R
+ \varsigma_{\ell}^{(a)} \,\bar \ell_L M_{\ell} \ell_R\right\}
\; +\;\mathrm{h.c.}
\eeqn

The flavour alignment results in a very specific structure, with all fermion-scalar
interactions being proportional to the corresponding fermion masses.
This leads to an interesting hierarchy of FCNC effects, suppressing them in light-quark
systems while allowing potentially relevant signals in heavy-quark transitions.
The only source of flavour-changing phenomena is the CKM matrix $V$, appearing in the
$W^\pm$ and $H_a^\pm$ interactions. Flavour mixing does not occur in the
lepton sector because of the absence of right-handed neutrinos.
The Yukawa Lagrangian is fully characterized in terms of the
$3 (N-1)$ complex parameters $\varsigma_{f}^{(a)}$ ($a\not= 1$), which
provide new sources of CP violation without tree-level FCNCs.

\subsection{The Aligned Two-Higgs-Doublet Model}

With $N=2$ one has the Aligned Two-Higgs-Doublet Model
(A2HDM) \cite{Pich:2009sp}, which contains one charged scalar field $H^\pm(x)$
and three neutral mass eigenstates  $\varphi^0_i(x) = \left\{h(x), H(x), A(x)\right\}$,
related through an orthogonal transformation
with the original fields $\mathcal{S}_i = \left\{S_1(x), S_2(x), P_2(x)\right\}$:
$\varphi^0_i(x) = \mathcal{R}_{ij} \,\mathcal{S}_j(x)$.
In the most general case, the CP-odd component $P_2$ mixes with the CP-even fields $S_{1,2}$ and the resulting scalar mass eigenstates do not have definite CP quantum numbers. For a CP-conserving scalar potential this admixture disappears, giving $A(x)= P_2(x)$ and
\bel{eq:A2HDM_Mixing}
\left(\! \ba h\\[3pt] H\ea\!\right)\, =\, \left[ \bat \cos{\tilde\alpha} & \sin{\tilde\alpha}\\[3pt]
- \sin{\tilde\alpha} & \cos{\tilde\alpha}\ea\right] \left( \!\ba S_1\\[3pt] S_2\ea\!\right)\, .
\ee
We adopt the conventions $M_h < M_H$ and $0\le \tilde\alpha \le\pi$.

The Yukawa Lagrangian is parametrized in terms of the three complex couplings $\varsigma^{(2)}_f\equiv \varsigma_f$, which encode all possible freedom allowed by the alignment conditions.
%Their flavour-blind phases provide an explicit counter-example to the
%widespread assumption that in two-Higgs-doublet models (2HDMs) without tree-level FCNCs all
%CP-violating phenomena should originate from the CKM matrix.
%
In terms of mass eigenstates,
\beqn\label{eq:a2hdm}
\cL_Y^{^{\mathrm{A2HDM}}}  & = &
-\frac{\sqrt{2}}{v}\; H^+ \left\{
\varsigma_{d} \,\bar u_L V M_{d} d_R
\, -\,\varsigma_{u} \,\bar u_R M_{u}^\dagger V d_L
\, + \,\varsigma_{\ell} \,\bar\nu_L  M_\ell\,\ell_R\right\}
\no\\ &&
- \sum_f\; \bar f_L M_f f_R\;\Bigl\{ 1 + \frac{1}{v}\;\sum_{\varphi^0_i} y_f^{\varphi^0_i}\;\varphi^0_i\Bigr\}
\, +\,\mathrm{h.c.}\, ,
\eeqn
with
\bel{eq:y_coup_A2HDM}
y_{d,\ell}^{\varphi^0_i}\, =\, \cR_{i1} + (\cR_{i2} + i\,\cR_{i3})\,\varsigma_{d,\ell}\, ,
\qquad\quad
y_u^{\varphi^0_i}\, =\, \cR_{i1} + (\cR_{i2} -i\,\cR_{i3}) \,\varsigma_{u}^*\, .
\ee
%

%%%%%%%%%%%%%%%%%%%%%% Higgs Signal Strengths %%%%%%%%%%%%%%%%%%%%%%%%%%%%%%
\begin{table}[tb]
\centering
\renewcommand{\arraystretch}{1.3} % enlarge line spacing
\renewcommand{\tabcolsep}{1pc} % enlarge column spacing
\begin{tabular}{lccc}
\hline
Model & $\varsigma_d$ & $\varsigma_u$ & $\varsigma_\ell$
\\ \hline
Type I & $\cot{\beta}$ & $\cot{\beta}$ & $\cot{\beta}$
\\
Type II & $-\tan{\beta}$ & $\cot{\beta}$ & $-\tan{\beta}$
\\
Type X & $\cot{\beta}$ & $\cot{\beta}$ & $-\tan{\beta}$
\\
Type Y & $-\tan{\beta}$ & $\cot{\beta}$ & $\cot{\beta}$
\\
Inert & 0 & 0 & 0
\\ \hline
\end{tabular}
\vskip .1cm
\caption{CP-conserving ${\cal Z}_2$ models ($\tan{\beta}\equiv v_2/v_1$) \cite{Pich:2009sp}.}
\label{tab:z2models}
\end{table}
%%%%%%%%%%%%%%%%%%%%%%%%%%%%%%%%%%%%%%%%%%%%%%%%%%%%%%%%%%%%%%%%%%%%%%%%%%%%%%%%%%%%%

The A2HDM constitutes a very general framework which includes, for particular values of its parameters, all previously considered two-Higgs doublet models without FCNCs  \cite{Branco:2011iw}.
FCNCs are usually avoided imposing appropriately chosen discrete $\cZ_2$ symmetries
such that only one scalar doublet couples to a given type of right-handed fermion field \cite{Glashow:1976nt,Paschos:1976ay}. Thus, one takes either $\cY^{(1)'}_f=0$ or $\cY^{(2)'}_f=0$ in Eq.~\eqn{eq:GenYukawa}. The choice can be different for $f=u,d,\ell$, leading to four different $\cZ_2$ models:
type I (all right-handed fermions couple to $\phi_2$) \cite{Haber:1978jt,Hall:1981bc}, type II ($d_R$ and $\ell_R$ couple to $\phi_1$, while $u_R$ couples to $\phi_2$) \cite{Hall:1981bc,Donoghue:1978cj}, type X (leptophilic or lepton specific; $\ell_R$ couples to $\phi_1$, and $d_R$ and $u_R$ couple to $\phi_2$) and type Y (flipped; $d_R$ couples to $\phi_1$, and $\ell_R$ and $u_R$ couple to $\phi_2$) \cite{Barger:1989fj}.
The resulting models are recovered for the particular values of $\varsigma_f$ indicated in Table~\ref{tab:z2models}. The $\cZ_2$ symmetries imply real alignment parameters and a CP-conserving scalar potential.

A different scenario appears if the $\cZ_2$ symmetry is imposed in the Higgs basis. In that case all right-handed fermions must couple to $\Phi_1$ in order to get their masses and, therefore, $\Phi_2$ decouples from the fermions and gauge bosons \cite{Deshpande:1977rw}.
The resulting Inert Doublet Model contains four dark scalars with limited
interactions with the SM particles. The lightest of them is stable and (if neutral) is a good candidate for DM. This inert model is in agreement with current data, both from accelerator and astrophysical experiments, and can accommodate the needed DM relic density. However, since the
CKM phase is the only source of CP violation, as in the SM, it fails to be a correct model for baryogenesis \cite{Krawczyk:2015xhl}.

The underlying discrete symmetry makes the flavour structure of the $\cZ_2$ models stable under quantum corrections (natural flavour conservation). This is no longer true in the more general A2HDM framework, where loop corrections generate a small misalignment of the Yukawa matrices
\cite{Pich:2009sp,Jung:2010ik,Ferreira:2010xe,Braeuninger:2010td}. However, the flavour symmetries of the A2HDM tightly constrain the possible FCNC structures, keeping their effects well below the present experimental bounds \cite{Pich:2009sp,Jung:2010ik,Li:2014fea,Abbas:2015cua}.

\subsection{A2HDM Phenomenology}

The built-in flavour symmetries protect very efficiently the aligned model from
unwanted effects, allowing it to easily satisfy the experimental constraints.
Leptonic and semileptonic decays are sensitive to tree-level $H^\pm$-exchange contributions
but, owing to the fermion-mass suppression of the Yukawa couplings, the resulting constraints
on the $\varsigma_f$ parameters are quite weak \cite{Jung:2010ik}.\footnote{
%%%%%%%%%%%%%%%%%%%%%
Some experimental flavour anomalies, such as the recently measured ratios $R(D^{(*)})\equiv \mathrm{Br}(B\to D^{(*)}\tau\nu)/\mathrm{Br}(B\to D^{(*)}\ell\nu)$ with $\ell=e,\mu$, are however difficult to accommodate within the A2HDM \cite{Celis:2012dk}.
}
%%%%%%%%%%%%%%%%%%%%%%
%
In spite of its flavour-blind CP phases,
the A2HDM satisfies also all present bounds on electric dipole moments \cite{Jung:2013hka}, although interesting signals could be expected within the projected sensitivity
of the next-generation of experiments.

\begin{figure}[bt]
\centering
\begin{minipage}[c]{5.5cm}\centering
\includegraphics[width=5.5cm]{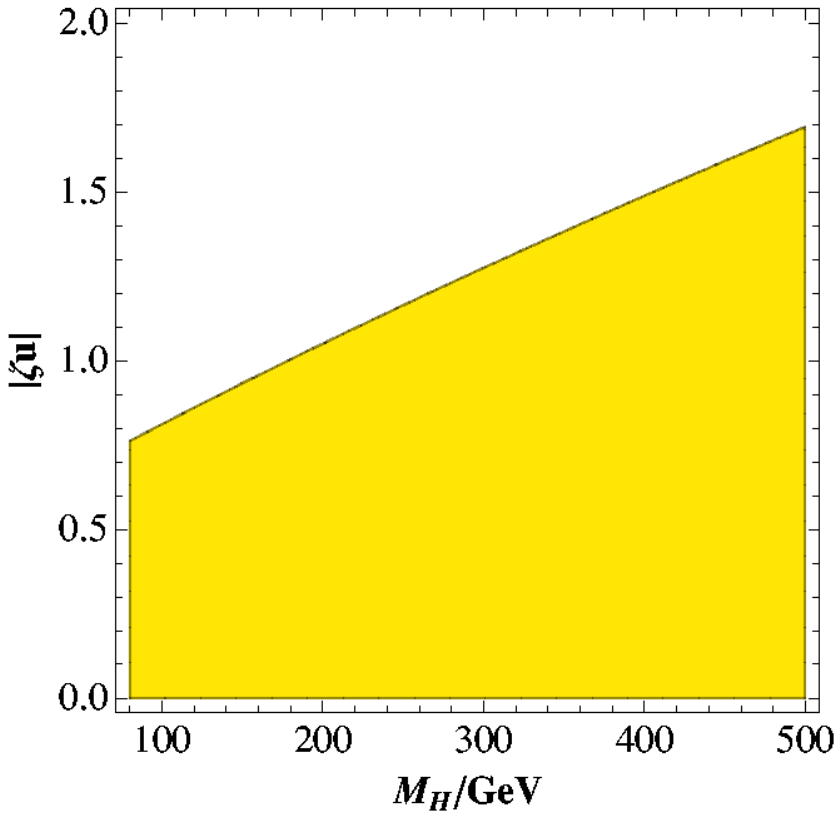}
\caption{95\% CL bounds on $\varsigma_u$, from $\epsilon_K$, as a function of $M_{H^\pm}$ \cite{Jung:2010ik}.}
\label{fig:EpsK}
\end{minipage}
\hfill
\begin{minipage}[c]{6.5cm}\centering
\includegraphics[width=6.5cm]{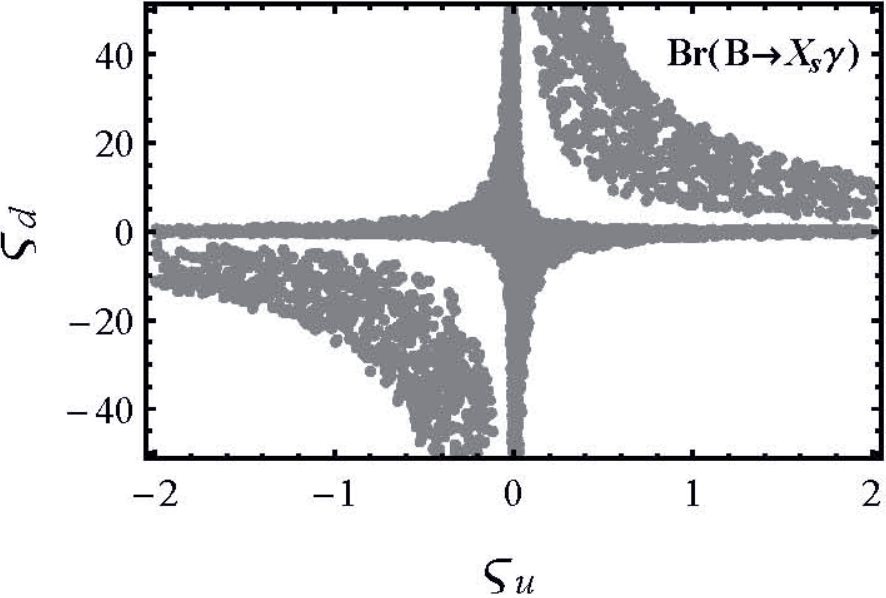}
\caption{95\% CL constraints on $\varsigma_{u,d}$ from $\bar B\to X_s\gamma$, assuming real couplings \cite{Jung:2012vu}.}
\label{fig:B2XsG}
\end{minipage}
\end{figure}

More stringent bounds are obtained from loop-induced transitions involving
virtual top-quark contributions, where the $H^\pm$ corrections are enhanced by the
top mass. Direct limits on $|\varsigma_u|$ can be derived from the $Z\to b\bar b$ decay width,
the mass difference $\Delta M_{B^0}$ between the $B^0$-$\bar B^0$ mass eigenstates,
%from $B^0$-$\bar B^0$ mixing
and the CP-violating parameter $\epsilon_K$ characterizing the mixing of neutral kaons.
The last observable provides the strongest limits, which are shown in Fig.~\ref{fig:EpsK}.
% as a function of $M_{H^\pm}$.
Other important constraints are obtained from rare FCNC decays such as $\bar B\to X_s\gamma$ \cite{Jung:2010ik,Jung:2010ab,Jung:2012vu} (Fig.~\ref{fig:B2XsG}) or
$B\to\ell^+\ell^-$ \cite{Li:2014fea}.

%%%%%%%%%%%%%%%%%%%%%%%%% Figures A2HDM %%%%%%%%%%%%%%%%%%%%%%%%
%%%%%%%%%%%%%%%%%%%%%%%%%%%%%%%%%%%%%%%%%%%%%%%%%%%%%%%%%%%%%%%%%%%%%%

The A2HDM leads also to a rich collider phenomenology. Neglecting CP-violation effects, the couplings of the neutral scalars to the fermions and gauge bosons are given,
in units of the SM Higgs couplings, by
\begin{align}  \label{equations1}
&& y_{f}^h & = \cos{\tilde\alpha} + \varsigma_f \sin{\tilde \alpha} \!\ , &&& y_{d,l}^A & =  i\,\varsigma_{d,l}  \!\ , &&  \notag \\
& & y_{f}^H & = -\sin{\tilde\alpha} + \varsigma_f \cos{\tilde \alpha} \!\ , &&&
y_{u}^A \; & =\; -i\, \varsigma_u  \!\  \,,
\end{align}
and ($\kappa_V^{\varphi_i^{0}}\equiv g_{\varphi_i^{0}VV}/g^{\mbox{\tiny SM}}_{HVV}$, $V=W^\pm,Z$)
\be\label{equations2}
\kappa_V^{h}\;=\; \cos{\tilde \alpha} \, , \qquad\qquad
\kappa_V^{H}\;=\; -\sin{\tilde \alpha} \, , \qquad\qquad
\kappa_V^{A}\;=\; 0  \, .
\ee
The CP symmetry implies a vanishing gauge coupling of the CP-odd scalar $A$.  In the limit $\tilde\alpha\to 0$, the $h$ couplings are identical to those of the SM Higgs field, the heavier CP-even scalar $H$ decouples from the gauge bosons and $\bigl|y_f^H\bigr| = \bigl|y_f^A\bigr| = \left|\varsigma_f\right|$. The opposite behaviour (up to a global sign) is obtained for $\tilde\alpha\to \pi/2$. The trivial trigonometric relation between $\sin{\tilde \alpha}$ and $\cos{\tilde \alpha}$ generates the sum rules \cite{Celis:2013ixa}
\bel{eq:sum-rules}
\bigl|\kappa_V^H\bigr|^2  = 1 -\bigl|\kappa_V^h\bigr|^2\, ,
\qquad\;
\bigl|y_f^H\bigr|^2 - \bigl|y_f^A\bigr|^2  = 1 - \bigl|y_f^h\bigr|^2\, ,
\qquad\;
\kappa_V^H \, y_f^H   = 1 -  \kappa_V^h \,  y_f^h \, ,
\ee
relating the couplings of the three neutral scalars.

Assuming that the lightest CP-even scalar $h$ is the Higgs boson discovered at 125~GeV, a global fit to the LHC data gives \cite{Celis:2013rcs,Celis:2013ixa}
\bel{eq:cos}
|\cos{\tilde \alpha}|\; >\;   0.90  \quad (0.80) \,,
\end{equation}
or equivalently $\sin{\tilde\alpha} <   0.44  \, \,(0.60)$, at 68\%~CL (90\%~CL).
%
%%%%%%%%%%%%%%%%%%%%%%%%%%%%%%%%%%%%%%%%%%%%%%%%%%%%%%%%%%%%%%%%%%%%%%%%%%
\begin{figure}[tb]
\centering
\includegraphics[width=6.cm,height=6.cm]{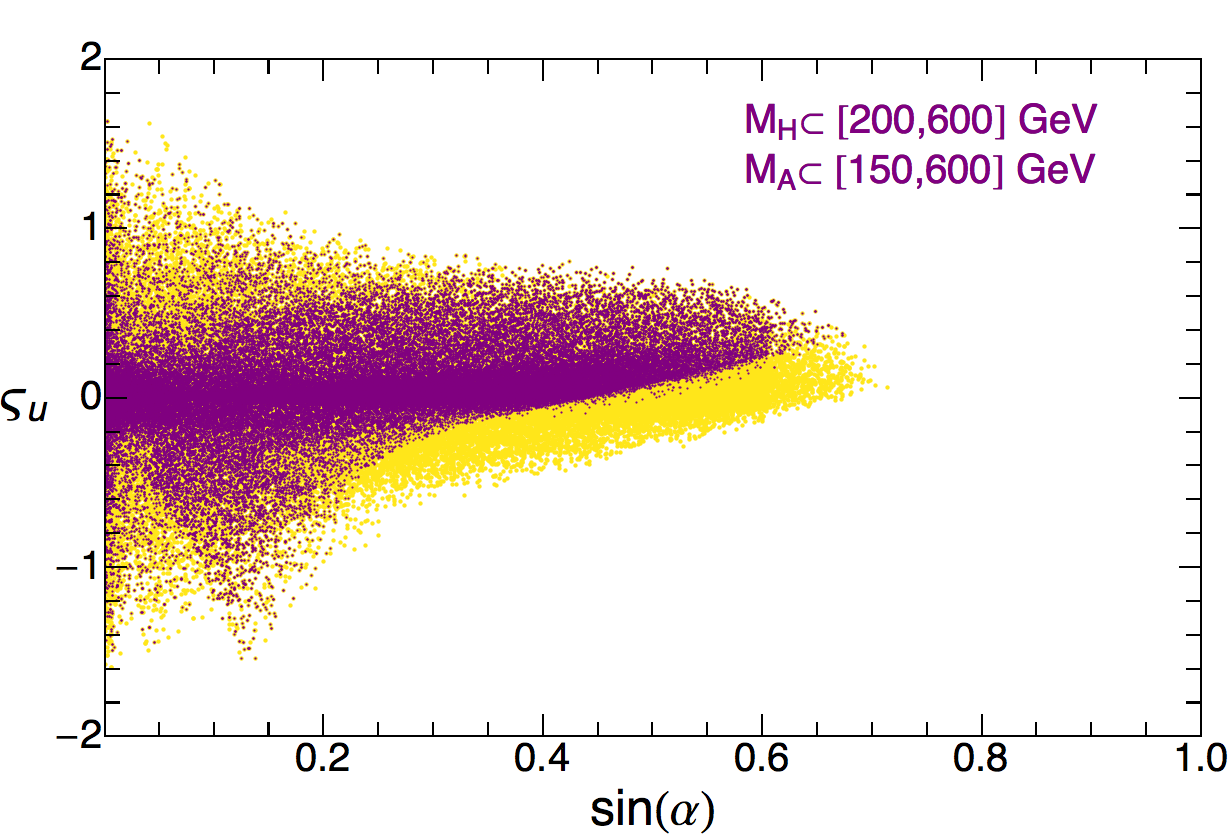}
\hfill
\includegraphics[width=6.cm,height=6.cm]{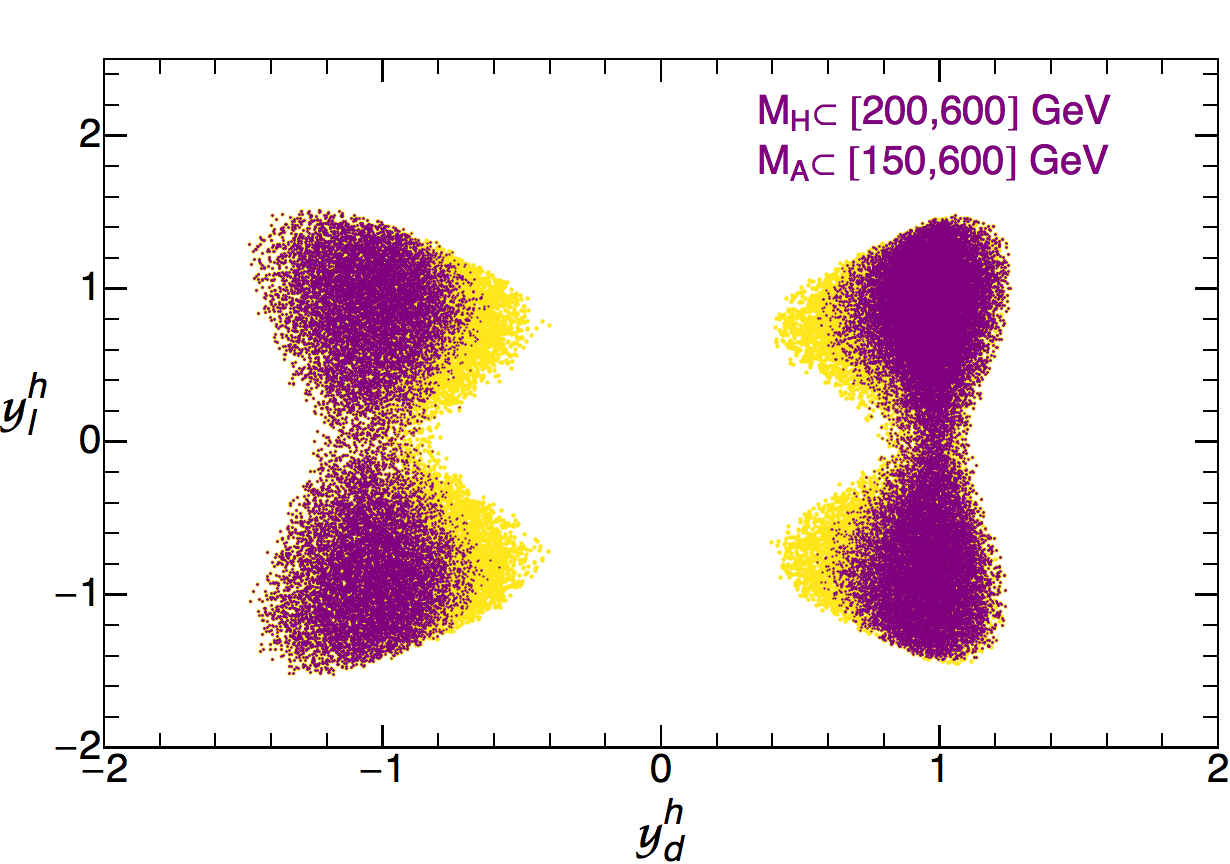}\\
\includegraphics[width=6.cm,height=6.cm]{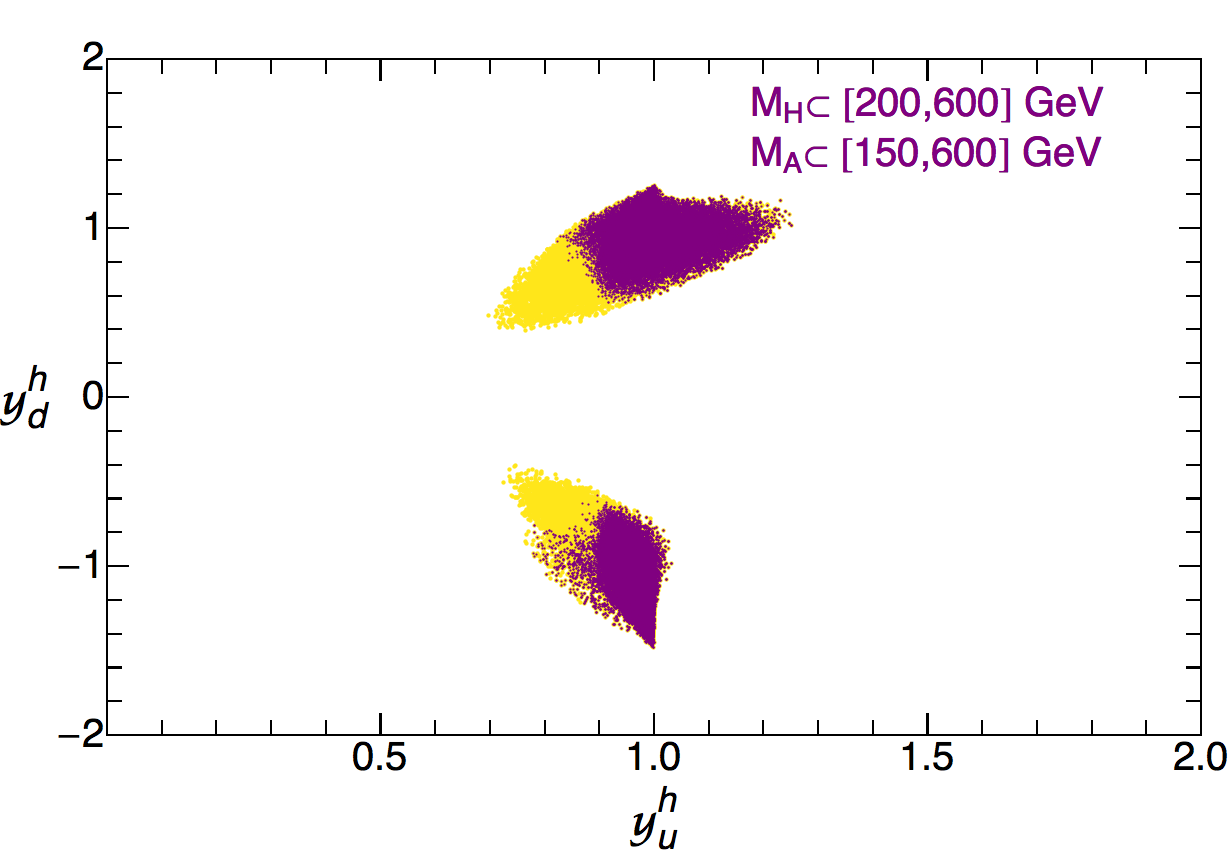}
\hfill
\includegraphics[width=6.cm,height=6.cm]{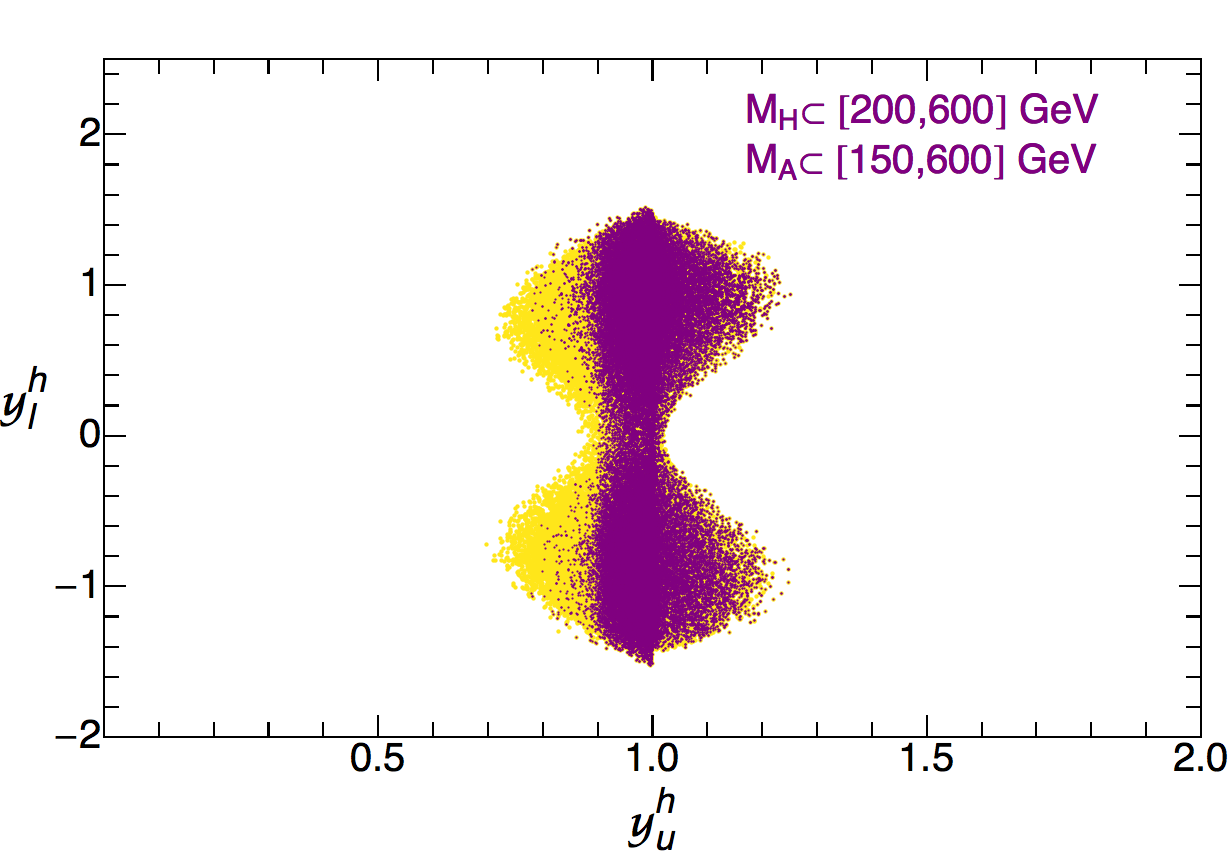}
\caption{Allowed 90\%~CL regions (yellow-light)
%in the planes  $\sin \tilde \alpha-\varsigma_u$ (top-left), $y_d^h- y_l^h$ (top-right),  $y_u^h- y_d^h$ (bottom-left), and $y^h_u - y_l^h$ (bottom-right),
from a global fit of $h(125)$ collider data together with $Z\to b\bar b$ and $\bar B \rightarrow X_s \gamma$, within the CP-conserving A2HDM. Neutral Higgs searches at the LHC, taking $M_{H} \in [200, 600]$~GeV and $M_{A} \in [150, 600]$~GeV, shrink the allowed regions to the purple-dark areas \cite{Celis:2013ixa}.}
\label{fig:RealiBII}
\end{figure}
%%%%%%%%%%%%%%%%%%%%%%%%%%%%%%%%%%%%%%%%%%%%%%%%%%%%%%%%%%%%%%%%%%%%%%%%%%
%
Figs.~\ref{fig:RealiBII}  \cite{Celis:2013ixa} show the allowed regions (yellow-light) at 90\%~CL, in the planes  $\sin \tilde \alpha-\varsigma_u$ (top-left), $y_d^h- y_l^h$ (top-right),  $y_u^h- y_d^h$ (bottom-left), and $y^h_u - y_l^h$ (bottom-right), including also in the fit the constraints from $Z\to b\bar b$ and $\bar B \rightarrow X_s \gamma$ data. The LHC searches for higher-mass neutral scalars provide complementary information, through the sum rules \eqn{eq:sum-rules}, shrinking the allowed regions to the purple-dark areas (assuming $M_{H} \in [200, 600]$~GeV and $M_{A} \in [150, 600]$~GeV).
Notice that the allowed parameter space would be much smaller in any of the ${\cal Z}_2$ models discussed before. For instance, in the type I model one has the additional restriction
$\varsigma_d = \varsigma_u =\varsigma_\ell$, which implies
$y_d^h = y_u^h = y_\ell^h$.

Although the gauge coupling of the $h(125)$ boson is found to be very close to the SM limit, there is still ample room for new physics effects related with the additional scalars. For instance, a very light fermiophobic charged Higgs ($\varsigma_f=0$), even below 80 GeV, is perfectly allowed by data \cite{Celis:2013ixa,Ilisie:2014hea}. The experimental bounds on new-physics contributions to the gauge-boson self-energies (Fig.~\ref{fig:GaugeSE} plus the additional diagrams with $H^\pm$) are satisfied, provided the mass differences $| M_{H^{\pm}} - M_H |$ and $| M_{H^{\pm}} - M_A |$ are not both large ($\gg 50$--100~GeV) at the same time \cite{Celis:2013rcs,Celis:2013ixa}. If a charged Higgs is found below the TeV scale, an additional neutral boson should be around.

\section{Custodial Symmetry and Dynamical EWSB}
%Strongly-Coupled Scenarios of EWSB}
\label{sec:custodial}

It is convenient to collect the SM Higgs doublet $\Phi$ and its charge-conjugate
$\tilde\Phi = i\sigma_2 \Phi^*$ into the $2\times 2$ matrix
\begin{equation}\label{eq:Sigma}
\Sigma\,\equiv\,\left(\tilde\Phi , \Phi\right)\, =\,
\left(\begin{array}{cc} \Phi^{0*} & \Phi^+ \\ -\Phi^- & \Phi^0\end{array}\right)
\, =\, \frac{1}{\sqrt{2}}\; (v+H)\; U(\vec\varphi)\, ,
\end{equation}
where the 3 Goldstone bosons are parametrized through the unitary matrix
\begin{equation}\label{eq:Uphi}
U(\vec\varphi)\;\equiv\;
\exp{\left\{\frac{i}{v}\:\vec{\sigma}\cdot\vec{\varphi}(x)\right\}}\, .
\end{equation}
With this field notation, the SM scalar Lagrangian (\ref{eq:Lphi}) adopts the form \cite{Pich:1998xt}:
\begin{eqnarray}\label{eq:Lsigma}\hskip -.2cm
\cL_\phi & =& \frac{1}{2}\;\mathrm{Tr} \left[ (D_\mu\Sigma)^\dagger D^\mu\Sigma\right]
- \frac{\lambda}{4}\,\left(\mathrm{Tr} \left[ \Sigma^\dagger \Sigma\right] - v^2\right)^2
\nonumber\\
& =& \frac{v^2}{4}\;\mathrm{Tr} \left[ (D_\mu U)^\dagger D^\mu U\right]
\; +\; \cO(H/v)\, .
\end{eqnarray}
This expression makes manifest the existence of a global
$SU(2)_L\otimes SU(2)_R$ symmetry:
\bel{eq:CustodialS}
\Sigma\,\to\, g_L^{\phantom{\dagger}}\,\Sigma\, g_R^\dagger\, ,
\qquad\qquad\qquad
g_X^{\phantom{\dagger}}\in SU(2)_X\, .
\ee
The ground state corresponds to a configuration proportional to the identity matrix, $\langle 0|\Sigma| 0\rangle = (v/\sqrt{2})\, I_2$, which is only preserved by those transformations satisfying $g_L=g_R$, {\it i.e.}, by the so-called custodial symmetry group
$SU(2)_{V}$ \cite{Sikivie:1980hm}. Thus, the SM scalar Lagrangian is characterized by the chiral symmetry breaking
\bel{eq:ChiralSymmetry}
SU(2)_L\otimes SU(2)_R\;\longrightarrow\; SU(2)_{V}\, .
\ee
The SM promotes the $SU(2)_L$ group to a local gauge symmetry, while only the $U(1)_Y$ subgroup of $SU(2)_R$ is gauged. The $SU(2)_R$ symmetry is then explicitly broken at $\cO(g')$ through the $U(1)_Y$ interaction in the covariant derivative.

The second line in (\ref{eq:Lsigma}), without the Higgs field, is the universal Goldstone Lagrangian associated with the chiral symmetry breaking \eqn{eq:ChiralSymmetry}. Any dynamical theory with this pattern of symmetry breaking has the same Goldstone interactions at energies much lower than the symmetry breaking scale $v$.
The same Lagrangian describes the low-energy chiral dynamics of the QCD pions, with the notational changes $v\to f_\pi$ and $\vec\varphi\to\vec\pi$ \cite{Pich:1998xt}.

The Goldstone covariant derivatives generate the masses of the gauge bosons. The successful relation between the $W^\pm$ and $Z$ masses is a consequence of the symmetry properties of
(\ref{eq:Lsigma}) and not of the particular dynamics implemented in the SM scalar potential. The gauge boson masses are not necessarily related to the Higgs field. The QCD pions also generate a tiny contribution to the $W^\pm$ and $Z$ masses, proportional to the pion decay constant $f_\pi$.

The EWSB could arise from some strongly-coupled underlying dynamics, similarly to what happens with chiral symmetry in QCD. The dynamics of the electroweak Goldstones and the Higgs boson can be analyzed in a model-independent way by using a low-energy effective Lagrangian \cite{Pich:1998xt},
based on the known pattern of chiral symmetry breaking in Eq.~\eqn{eq:ChiralSymmetry}.
%$\mathrm{SU(2)}_L\otimes \mathrm{SU(2)}_R\to \mathrm{SU(2)}_{L+R}$.
In full generality, the Higgs is taken as a singlet field, unrelated to the Goldstone triplet $\vec\varphi$.
The effective Lagrangian is organized as a low-energy expansion in powers of momenta (derivatives) and symmetry breakings:
%$\cL = \sum_n \cL_n$.
%
\bel{eq:Leff}
\cL_{\mathrm{eff}} \, =\,  \sum_{n=2} \cL_n\, .
\ee
At lowest order (LO), $\cO(p^2)$, it contains the renormalizable massless (unbroken) SM Lagrangian, plus the Goldstone term in (\ref{eq:Lsigma}) multiplied by an arbitrary function of the Higgs field\footnote{
%%%%%%%%%%%%%%%%%%%%%%%%%%%%%%%%%%%%%%%%%%%%
The non-linear representation of the Goldstone fields in Eq.~\eqn{eq:Uphi} contains arbitrary powers of $\vec\varphi$ compensated by corresponding powers of the electroweak scale $v$. If $H$ and $\vec\varphi$ are assumed to have similar origins, powers of $H/v$ do not increase either the chiral dimension.}
%%%%%%%%%%%%%%%%%%%%%%%%%%%%%%%%%%%%%%%%%%%%
$\cF_u(H/v) = \sum_{n=0} \cF_{u,n}\; (H/v)^n$ \cite{Grinstein:2007iv}.
The Lagrangian $\cL_2$ includes in addition the kinetic Higgs Lagrangian and a scalar potential $V(H/v)$ containing arbitrary powers of $H/v$.

\begin{figure}[t]
\centering
\includegraphics[width=11cm,clip]{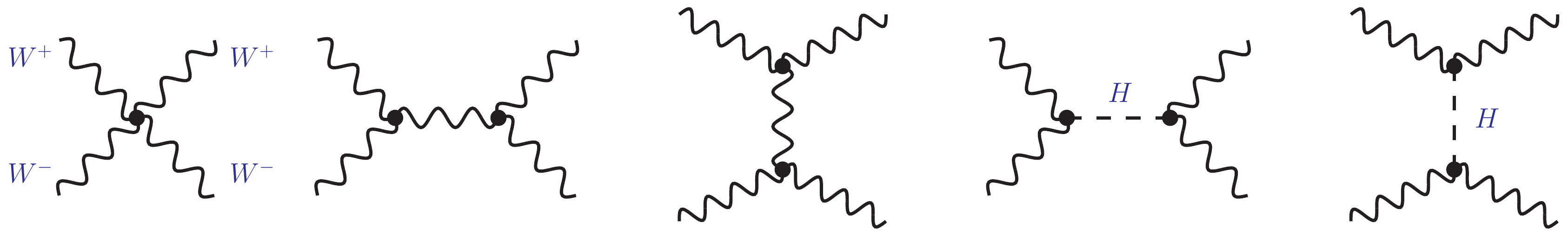}
\caption{Tree-level contributions to the scattering of gauge bosons.}
\label{fig:WWscattering}
\end{figure}

Since the electroweak Goldstones constitute the longitudinal polarizations of the gauge bosons, the scattering $V_L V_L\to V_L V_L$ ($V=W^\pm, Z$) directly tests the Goldstone dynamics. In the absence of the Higgs field, the tree-level scattering matrix has the same form as the elastic scattering amplitude of QCD pions:
\beqn\label{eq:pi-pi}
T(W^+_L W^-_L\to W^+_L W^-_L) & = & T(\varphi^+\varphi^-\to \varphi^+\varphi^-)\, +\, \cO\left(\frac{M_W}{\sqrt{s}}\right)
\no\\
& = & \frac{s+t}{v^2}\, +\, \cO\left(\frac{M_W}{\sqrt{s}}\right)\, .
\eeqn
At high energies it grows as $s/v^2$ which implies an unacceptable violation of unitarity. In the SM, the right high-energy behaviour is recovered through the additional contributions from Higgs boson exchange in Fig.~\ref{fig:WWscattering}, which exactly cancel the unphysical growing:
\beqn
T_{\mbox{\tiny SM}}& =&\frac{1}{v^2}\,\left\{ s + t - \frac{s^2}{s-M_H^2}
- \frac{t^2}{t-M_H^2}\right\}\, =\,
-\frac{M_H^2}{v^2}\,\left\{ \frac{s}{s-M_H^2}+ \frac{t}{t-M_H^2}\right\} .
\no\\ &&
\eeqn
However, this subtle cancellation is destroyed with generic $HV^2$ %and $H^2V^2$
couplings.
One-loop corrections induce a much worse ultraviolet behaviour $s^2\log{s}/v^4$ \cite{Delgado:2013hxa,Espriu:2013fia}, which is only cancelled if the gauge couplings of the Higgs boson take exactly their SM values. Any small deviation from the SM in the Higgs couplings would necessarily imply the presence of new-physics contributions to the $V_L V_L\to V_L V_L$ scattering amplitude, in order to restore unitarity. The same applies to the gauge-boson self-interactions which are also relevant in this unitarity cancellation.
Therefore, the measurement of $\sigma(V_L V_L\to V_L V_L)$ at the LHC is a very important, but difficult, challenge.
The first evidence of $W^\pm W^\pm$ collisions has been recently reported by ATLAS \cite{Aad:2014zda}, and used to set limits on anomalous quartic gauge couplings.

%%%%%%%%%%%%%%%%%%%%%%% Figure xSM %%%%%%%%%%%%%%%%%%%%%
%%%%%%%%%%%%%%%%%%%%%%%%%%%%%%%%%%%%%%%%%%%%%%%%%%%%%%%%%%%%%%%%%%%%
%

At the next-to-leading order, one must consider one-loop contributions from the LO Lagrangian plus $\cO(p^4)$ local structures:
\be
\cL_4 \, =\, \sum_i \cF_i(H/v)\; \cO_i\, ,
\qquad\qquad
\cF_i(H/v)\, =\, \sum_{n=0} \cF_{i,n}\; \left(\frac{H}{v}\right)^n\, .
\ee
Writing the most general basis of operators $\cO_i$
\cite{Longhitano:1980tm,Longhitano:1980iz,Appelquist:1980vg,Buchalla:2012qq,Alonso:2012px,Buchalla:2013rka,Pich:2015kwa}, build with the known (light) fields and the SM symmetries, the effective Lagrangian parametrizes the low-energy effects of any underlying short-distance theory compatible with these symmetries.

In the absence of direct discoveries of new particles, the only possible signals of the high-energy dynamics are hidden in the couplings of the low-energy electroweak effective theory
\cite{Pich:2015kwa}, which can be tested through scattering amplitudes among the known particles. They can be accessed experimentally through precision measurements of anomalous triple and quartic gauge couplings, scattering amplitudes of longitudinal gauge bosons, Higgs couplings, etc. \cite{Schladming}.

\section{Status and Outlook}
\label{sec:outlook}

After the Higgs discovery, the SM framework is now fully established as the correct theory describing the interactions of the elementary particles at the electroweak scale.
%It successfully explains the experimental results with high precision and all its ingredients have been verified.
With the measured Higgs and top masses, the SM could even be a valid theory up to the Planck scale.
However, new physics is still needed to explain many pending questions for which we are lacking a proper understanding,
such as the matter-antimatter asymmetry, the pattern of flavour mixings and fermion masses, the nature of dark matter or the accelerated expansion of the Universe.
The SM accommodates the measured masses, but it does not explain the vastly different mass scales spanned by the known particles. The dynamics of flavour and the origin of CP violation are also related to the mass generation.

The Higgs boson could well be a window into unknown dynamical territory, may be also related to the intriguing existence of massive dark objects in our Universe. Therefore, the Higgs properties must be analyzed with high precision to uncover any possible deviation from the SM. The present data are already putting stringent constraints on alternative scenarios of EWSB and pushing the scale of new physics to higher energies. How far this scale could be is an open question of obvious experimental relevance.

The ongoing LHC run could bring interesting surprises. The first results released from the
full 2015 data sample, at $\sqrt{s} = 13$~TeV, show already some hints of a possible $2\gamma$
resonance structure at $M_{2\gamma} = 750$~GeV \cite{ATLAS-CONF-2015-081,CMS:2015dxe,CERN}. The statistical significance is still low and the signal could well be just a statistical fluctuation; otherwise we could be witnessing the first indications of a second scalar boson and the emergence of a new fundamental paradigm. In any case, there is no doubt that as more data will get accumulated we will learn which directions Nature has chosen to organize the microscopic world of particle physics. We are awaiting for great discoveries; the LHC scientific adventure is just starting.

\section*{Acknowledgments}
I would like to thank the organizers for inviting me to present these lectures and all the School participants for their many interesting questions.
This work has been supported by the Spanish Government and ERDF funds from the European Commission [FPA2011-23778, FPA2014-53631-C2-1-P], by the Spanish Centro de Excelencia
Severo Ochoa Programme [SEV-2014-0398] and by the Generalitat Valenciana [PrometeoII/2013/007].

\end{document}